\documentclass{aa}

\usepackage{graphicx}
\usepackage{amsmath}
\usepackage{amssymb}
\usepackage{multicol}
\usepackage{bm}
\usepackage{pdflscape}
\usepackage{graphicx}
\usepackage{xcolor}
\usepackage{subfig}
\usepackage{caption}

\usepackage[varg]{txfonts}

\usepackage[colorlinks=true,citecolor=blue]{hyperref}

\usepackage[T1]{fontenc}
\usepackage{ae,aecompl}

\usepackage{natbib}
\bibpunct{(}{)}{;}{a}{}{,}

\newcommand{\sgas}{SGAS\,J0033$+$02}
\newcommand{\sgaslong}{SGAS\,J003341.5$+$024217}

\newcommand{\oii}{[\ion{O}{ii}]}
\newcommand{\mgii}{\ion{Mg}{ii}}
\newcommand{\mgi}{\ion{Mg}{i}}
\newcommand{\feii}{\ion{Fe}{ii}}
\newcommand{\caii}{\ion{Ca}{ii}}
\newcommand{\mnii}{\ion{Mn}{ii}}
\newcommand{\nai}{\ion{Na}{i}}
\newcommand{\hi}{\ion{H}{i}}
\newcommand{\civ}{\ion{C}{iv}}
\newcommand{\ciii}{\ion{C}{iii}]}

\newcommand{\kms}{km\,s$^{-1}$}

\newcommand{\sfroii}{$70.0\pm 1.6$}
\newcommand{\sfrsed}{$63.7\pm 0.1$}

\newcommand{\arctomo}{{\sc Arctomo}}
\newcommand{\megasaura}{{\sc Meg}a{\sc S}a{\sc ura}}
\newcommand{\hst}{{\it HST}}
\newcommand{\lenstool}{\tt{Lenstool}\rm}
\newcommand{\bagpipes}{\tt{Bagpipes}\rm}
\newcommand{\pyspeckit}{\tt{PySpecKit}\rm}
\newcommand{\mpdaf}{\tt{MPDAF}\rm}

\newcommand{\pymuse}{\tt{PyMUSE}\rm}

\newcommand{\galpak}{$\mathrm{GalPak}^\mathrm{3D}$}

\newcommand{\eso}{European Southern Observatory, Alonso de C\'ordova 3107, Vitacura, Casilla 19001, Santiago, Chile\label{eso}}
\newcommand{\puc}{Instituto de Astrof\'\i sica, Pontificia Universidad Cat\'olica de Chile, Casilla 306, Santiago, Chile\label{puc}}

\newcommand{\pucv}{Instituto de F\'\i sica, Pontificia Universidad Cat\'olica de Valpara\'\i so, Casilla 4059, Valpara\'\i so, Chile\label{pucv}}
\newcommand{\vic}{Department of Physics and Astronomy, Camosun College, 3100 Foul Bay Road, Victoria, BC, Canada V8P 5J2\label{vic}}
\newcommand{\cral}{Centre de Recherche Astrophysique de Lyon, UMR5574, 9 avenue Charles Andr\'e, 69230 Saint-Genis-Laval, France\label{cral}}
\newcommand{\udp}{Instituto de Estudios Astrof\'\i sicos, Universidad Diego Portales, Av. Ej\'ercito Libertador 441, Santiago, Chile\label{udp}}
\newcommand{\liege}{STAR Institute, Quartier Agora, All\'ee du Six Ao\^ut, 19c B-4000 Li\`ege, Belgium\label{liege}}
\newcommand{\iap}{Institut d'Astrophysique de Paris, UMR7095, 98bis boulevard Arago, 75014 Paris, France\label{iap}}
\newcommand{\fcla}{French-Chilean Laboratory for Astronomy, IRL3386, CNRS and U. de Chile, Casilla 36-D, Santiago, Chile\label{fcla}}

\begin{document}

\title{Unveiling large-scale rotational motions in the intragroup medium at $z\sim 1$
through gravitational-arc
tomography\thanks{Based on observations carried out in service mode at the European
Southern Observatory (ESO) under programme ID 098.A-0459(A).}}

\titlerunning{Kinematics of the IGrM at $z\sim 1$}
\authorrunning{C. Ledoux et al.}

\author{C\'edric~Ledoux\inst{\ref{eso},}\corrauth{cledoux@eso.org}
\and Fernanda~Mu\~noz-Olivares\inst{\ref{eso},\ref{puc}}
\and L.~Felipe~Barrientos\inst{\ref{puc}}
\and Nicolas~Tejos\inst{\ref{pucv}}
\and Trystyn~Berg\inst{\ref{vic}}
\and Felipe~Corro-Guerra\inst{\ref{cral}}
\and Evelyn~Johnston\inst{\ref{udp}}
\and Guillaume~Mahler\inst{\ref{liege}}
\and Jorge~Gonz\'alez-L\'opez\inst{\ref{puc}}
\and Joaqu\'\i n~Hern\'andez-Guajardo\inst{\ref{puc}}
\and Pasquier~Noterdaeme\inst{\ref{iap},\ref{fcla}}}

\institute{\eso \and \puc \and \pucv \and \vic \and \cral \and \udp \and \liege \and \iap \and \fcla}

\date{Received 27 February 2026 / Accepted 12 March 2026}

\abstract
{The circumgalactic medium (CGM) is a crucial interface between galaxies and their large-scale
environment, regulating gas accretion and feedback processes. Yet, its physical and kinematic
properties within galaxy groups, where most galaxies reside, remain poorly constrained.}
{We present the first spatially resolved characterisation of the cool intragroup medium (IGrM) in
a spectroscopically confirmed galaxy group at $z\simeq 1.167$, using absorption-line spectroscopy
along multiple sightlines.}
{Using 30 independent sightlines towards the gravitationally lensed galaxy \sgaslong, we combine
background light from an extended gravitational arc and various sources in the field to
map the distribution and kinematics of diffuse, metal-enriched gas pertaining to this group.}
{We detect prominent \mgii, \feii, \caii, and \mgi\ absorption extending up to a projected
distance of 62~kpc from a massive ($\log M_\star=11.0\pm 0.1$~M$_\sun$) star-forming spiral and
its interacting companion. Together with four other members, these form a compact group with
a virial radius of $313$~kpc. Down-the-barrel, blueshifted absorption indicates outflows. The
distribution and two-dimensional kinematics of this gas suggest the influence of both tidal
stripping and star formation-driven winds. Intervening absorption across the field partly
traces internal galaxy motions. A simple superposition of individual discs cannot reproduce
the velocity field at large impact parameters or in counter-rotating regions, while a
global IGrM halo with a rotational velocity of $\approx 130$~km~s$^{-1}$ provides a good
match. Beyond individual galaxy envelopes, the data are found to be consistent with a group-scale
structure that co-rotates in concert with the galaxies. Assuming dynamical equilibrium,
we estimate a total (cool$+$warm$+$hot) gas mass of $1.3-2.5\times 10^{11}$~$M_\odot$, with
large systematic uncertainties, corresponding to roughly 50\% of all baryons, within
one-quarter of the group's virial radius.}
{These results point to a multiphase IGrM in which cool ($\sim 10^4$~K) clouds are embedded
within a dynamically coherent, group-wide halo. The gas appears gravitationally bound to the
group rather than reaccreting onto individual galaxies. High-redshift strong \mgii\
absorbers may thus trace shared, metal-enriched halos shaped by galaxy interactions
and feedback, with stripped and outflowing gas accumulating in the IGrM over time.}

\keywords{quasars: absorption lines -- galaxies: halos -- galaxies: interactions -- galaxies:
clusters: individual (\sgaslong)}

\maketitle
\nolinenumbers

\section{Introduction}

The diffuse gas component of the Universe plays a central role in the formation of galaxies
and drives their evolution across cosmic time. In this context, the circumgalactic medium (CGM)
serves as a transition region between the pervasive cosmic web on megaparsec scales and
the inner star-forming regions of galaxies \citep{Tumlinson17,Peroux20}. Numerical
simulations predict that cool gas streams from the large-scale structures become
collimated and contribute to generating angular momentum within the CGM, a phenomenon
still poorly constrained observationally \citep[e.g.,][]{DeFelippis20,DeFelippis21}.
In addition, metal-enriched gas and dust produced by star formation can be expelled
from galaxies to the CGM through powerful stellar winds and active galactic nuclei (AGN),
as well as processes such as tidal and ram-pressure stripping. This leads to a complex
interplay between various gaseous phases \citep[see][]{Faucher-Giguere23}.

Diffuse gas can be studied through absorption lines detected against background sources such as
quasars and $\gamma$-ray burst afterglows. Modern spectroscopic quasar surveys
routinely identify systems with large neutral-hydrogen column densities at intermediate and
high redshifts, known as damped Lyman-$\alpha$ absorbers (DLAs), as well as
strong \mgii\ systems \citep[e.g.,][]{Noterdaeme09,Chabanier22}. These absorbers are believed
to originate from gas associated with galaxies, as indicated by their significant metal content.
DLAs usually consist of a mixture of warm and cool ($\sim 10^4$~K) gas with low
average molecular fractions \citep{Ledoux03,Noterdaeme08}. Due to these characteristics and
the preferential selection of absorbers at large galactocentric distances, the bulk of the gas
traced by metal-rich quasar absorbers is more likely located in the outskirts of galaxies than
in the most central regions of star formation \citep[e.g.,][]{Krogager12,Muzahid15,Neeleman25}.

To date, cool gas in the CGM at $z\sim 1$ has predominantly been studied using \mgii\ absorption
by leveraging the statistical power of numerous, randomly distributed quasar sightlines. For
example, \citet{Huang21} showed that the rest-frame equivalent width of the \mgii\,$\lambda$2796
line depends, on average, on the halo radius, B-band luminosity, and stellar mass of the
galaxies associated with the absorption. The covering fraction of \mgii-bearing gas is
typically high, exceeding 60\% within 40~kpc of isolated galaxies, and it declines
rapidly, vanishing beyond projected distances of $\gtrsim 100$~kpc. However, while
quasar sightlines provide valuable insights, they remain sparse and usually offer a single pencil
beam per field \citep[e.g.,][]{Weng23}. As such, complementary constraints from individual halos
are needed to fully understand the properties of the CGM.

To investigate the CGM of individual galaxies or galaxies in groups, \citet{Lopez18} pioneered
a technique known as gravitational-arc tomography (hereafter referred to as \arctomo).
This approach utilises integral-field spectroscopic observations of the brightest
known extended gravitational arcs in the sky, providing multiple, contiguous or closely
spaced, sightlines within each field. This represents an advancement in CGM studies, as it
enables detailed examinations of gas distributions around galaxies and spatially
resolved measurements of velocity fields within individual halos
\citep{Lopez20,Mortensen21,Fernandez-Figueroa22,Bordoloi22}. To date, such studies have revealed
a complex picture, in which metal-enriched gas exhibits clumpy
distributions \citep[e.g.,][]{Afruni23} and/or co-rotates with stellar discs out to distances of
tens of kiloparsecs \citep{Lopez20,Tejos21}. Additionally, observations of outflows provide
insights into metal recycling processes within the CGM \citep[e.g.,][]{Nielsen20,Tejos21}.

In this paper, we use \arctomo\ to explore the kinematics of DLA-like gas in the vicinity of
galaxies in groups, where dynamical effects are expected, extending our analysis of the CGM
to the IGrM. We present Very Large Telescope (VLT) observations of the bright lensed galaxy
\sgaslong\ (hereafter \sgas), integrating emission and absorption line data to constrain
both the distribution and kinematics of diffuse, metal-enriched gas traced by \mgii\
absorption, within a galaxy group at $z\simeq 1.17$.
Throughout this work, we adopt a spatially flat $\Lambda$CDM cosmology with
$\Omega_\mathrm{m}=0.3$ and $H_0=70$~km~s$^{-1}$~Mpc$^{-1}$.

\section{Observations}\label{sec:observations}

\subsection{Target selection and {\it HST}/WFC3 imaging}

The bright lensed galaxy \sgas\ at $z\simeq 2.39$ was discovered during a survey aimed at
identifying giant, bright gravitational arcs, which contributed to creating the
Magellan Evolution of Galaxies Spectroscopic and Ultraviolet Reference Atlas, known
as \megasaura\ \citep{Rigby18a}. In the spectrum of this source, strong \mgii\ absorption
lines were detected at $z\simeq 1.17$ within the arc continuum emission, leading us to
identify this field as a target for gravitational-arc tomography.

The field of \sgas\ was previously observed with the Wide Field Camera~3 onboard \hst\
during two visits on October 30 and November 8, 2016 (proposal ID 14170; PI: Wuyts). In
the IR channel, images were captured using the F105W and F140W filters, while in the UVIS
channel, exposures were taken with the F410M, F555W, and F814W filters. The
observations and data reduction procedures are described in \citet{Fischer19}.

\begin{figure*}
\sidecaption
\includegraphics[width=0.67\hsize]{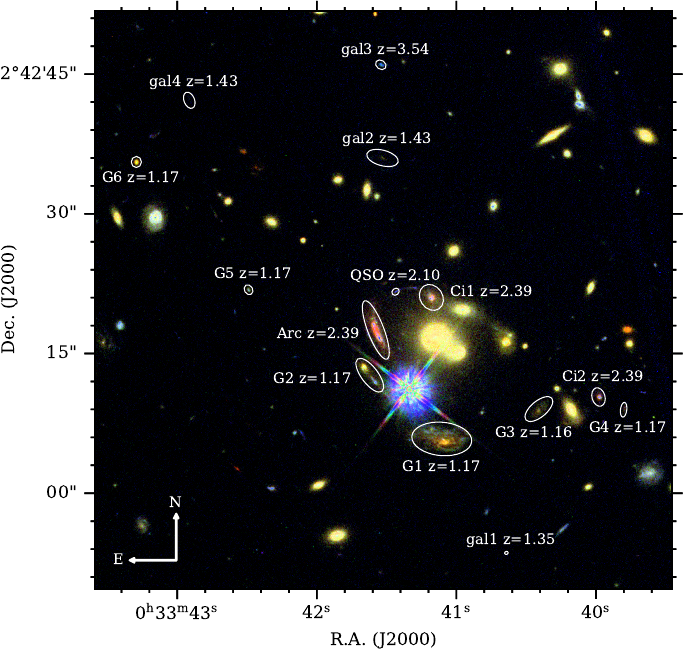}
\caption{Field of \sgas\ observed with \hst\ in the optical (F814W) and near-infrared (F105W and F140W) bands.
The gravitational arc formed by the lensed galaxy at $z\simeq 2.39$ (called the 'Arc') and its
counter-images ('Ci1' and 'Ci2') are indicated. The bluish object located $7\farcs 3$ southwest of
the arc is a Galactic star. The main absorbing galaxy at $z\simeq 1.17$, labelled 'G1', is shown
along with its closest two companions at the same redshift: 'G2' to the northeast, and 'G3' to the
west. Redshifts for these and additional sources at $z>1.16$ are indicated next to each label.}
\label{fig:field}
\end{figure*}

Fig.~\ref{fig:field} shows a multi-colour \hst\ composite image of the field, spanning the
optical to near-infrared range.
The lensing model presented in \citet{Fischer19} revealed a system comprising one main
gravitational arc and two counter-images of the lensed galaxy. The foreground lensing, massive
galaxy cluster is located at $z=0.4716$. Additionally, ALMA 870~$\mu$m observations
by \citet{Solimano21} detected dust-continuum emission from a prominent galaxy in the field,
which we will refer to as ``G1''.

\begin{table*}
\caption{Census of extragalactic sources at $z>1.16$.}\label{tab:gal}
\centering
\begin{tabular}{lllcllll}
\hline\hline
Object & MARZ classification & $z$\tablefootmark{a} & Lines
       & R.A.\tablefootmark{b} & Dec.\tablefootmark{b} & $\theta$\tablefootmark{b,c} & \mgii\ at                 \\
       &                     &                      & used
       & (J2000)               & (J2000)               & [\arcsec]                   & $z_{\rm abs}\simeq 1.17$? \\
\hline
\multicolumn{2}{l}{Group galaxies at $z\simeq 1.17$:} &&&&&&\\
G1     & Late type emission galaxy    & 1.1672(3) & \oii  & $0^{\rm h}$ $33^{\rm m}$ $41\fs 10$ & $+2\degr$ $42\arcmin$  $5\farcs 6$
       &   0.0  & Yes\tablefootmark{d} \\
G2     & Star-forming galaxy          & 1.1674(0) & \oii  & $0^{\rm h}$ $33^{\rm m}$ $41\fs 61$ & $+2\degr$ $42\arcmin$ $12\farcs 7$
       &  10.43 & Yes\tablefootmark{d} \\
G3     & Late type emission galaxy    & 1.1649(9) & \oii  & $0^{\rm h}$ $33^{\rm m}$ $40\fs 40$ & $+2\degr$ $42\arcmin$  $9\farcs 1$
       &  11.06 & Yes\tablefootmark{d} \\
G4     & Star-forming galaxy          & 1.1667(6) & \oii  & $0^{\rm h}$ $33^{\rm m}$ $39\fs 79$ & $+2\degr$ $42\arcmin$  $9\farcs 1$
       &  19.94 & ...\tablefootmark{e} \\
G5     & Late type emission galaxy    & 1.1674(5) & \oii  & $0^{\rm h}$ $33^{\rm m}$ $42\fs 48$ & $+2\degr$ $42\arcmin$ $22\farcs 0$
       &  26.39 & No                   \\
G6     & Transitional galaxy          & 1.1673(8) & \oii  & $0^{\rm h}$ $33^{\rm m}$ $43\fs 28$ & $+2\degr$ $42\arcmin$ $35\farcs 7$
       &  44.42 & ...\tablefootmark{e} \\
\hline
\multicolumn{2}{l}{Lensed galaxy \sgas:} &&&&&&\\
Arc    & High-$z$ star-forming galaxy & 2.3874  & \ciii & $0^{\rm h}$ $33^{\rm m}$ $41\fs 57$ & $+2\degr$ $42\arcmin$ $17\farcs 6$
       &  13.91 & Yes                  \\
Ci1    & High-$z$ star-forming galaxy & 2.3874  & \ciii & $0^{\rm h}$ $33^{\rm m}$ $41\fs 17$ & $+2\degr$ $42\arcmin$ $21\farcs 1$
       &  15.54 & Yes                  \\
Ci2    & High-$z$ star-forming galaxy & 2.3874  & \ciii & $0^{\rm h}$ $33^{\rm m}$ $39\fs 97$ & $+2\degr$ $42\arcmin$ $10\farcs 5$
       &  17.63 & Yes                  \\
\hline
\multicolumn{2}{l}{Additional sources:} &&&&&&\\
gal1   & Transitional galaxy          & 1.3510(7) & \oii  & $0^{\rm h}$ $33^{\rm m}$ $40\fs 63$ & $+2\degr$ $41\arcmin$ $53\farcs 7$
       &  13.83 & ...\tablefootmark{e} \\
QSO    & Quasar                       & 2.0966    & \ciii & $0^{\rm h}$ $33^{\rm m}$ $41\fs 43$ & $+2\degr$ $42\arcmin$ $21\farcs 7$
       &  16.84 & Yes                  \\
gal2   & Transitional galaxy          & 1.4300(3) & \oii  & $0^{\rm h}$ $33^{\rm m}$ $41\fs 52$ & $+2\degr$ $42\arcmin$ $36\farcs 1$
       &  31.14 & No                   \\
gal3   & High-redshift galaxy         & 3.5421:   & \hi, \civ\tablefootmark{f} & $0^{\rm h}$ $33^{\rm m}$ $41\fs 53$ & $+2\degr$ $42\arcmin$ $46\farcs 1$
       &  41.01 & No                   \\
gal4   & Transitional galaxy          & 1.4308(3) & \oii  & $0^{\rm h}$ $33^{\rm m}$ $42\fs 90$ & $+2\degr$ $42\arcmin$ $42\farcs 3$
       &  45.54 & ...\tablefootmark{e} \\
gal5   & Star-forming galaxy          & 1.4312(8) & \oii  & $0^{\rm h}$ $33^{\rm m}$ $43\fs 66$ & $+2\degr$ $42\arcmin$ $52\farcs 4$
       &  60.51 & ...\tablefootmark{e} \\
\hline
\end{tabular}
\tablefoot{
\tablefoottext{a}{Redshift derived from integrated spectra;}
\tablefoottext{b}{Image plane;}
\tablefoottext{c}{Angular distance from G1;}
\tablefoottext{d}{Down-the-barrel absorption;}
\tablefoottext{e}{Faint continuum;}
\tablefoottext{f}{Absorption lines.}}
\end{table*}

\subsection{VLT/MUSE integral-field spectroscopy}\label{sec:data}

Follow-up observations of \sgas\ were conducted at the VLT as part of ESO programme ID
098.A-0459(A) (PI: Lopez) using the Multi-Unit Spectroscopic Explorer \citep[MUSE;][]{Bacon10}.
MUSE is an optical integral-field spectrograph with a $1\arcmin\times 1\arcmin$ field of
view in its Wide-Field mode and a native spatial (spaxel) scale of $0\farcs2$. The instrument
yields a resolving power of \textit{R}$\sim 1770$ at 480\,nm and \textit{R}$\sim 3590$ at
930\,nm.

The data were acquired in natural-seeing mode during dark time on the nights of September
19 and 20, 2017. Throughout the observations, the seeing conditions at the zenith remained 
stable at $0\farcs 6$ FWHM, as measured by the Differential Image Motion Monitor (DIMM)
in V-band. The sky transparency conditions were photometric, and the target airmass ranged
from 1.15 to 1.5 during the observations. A total of eleven exposures were obtained,
each with an integration time of 701\,s, distributed across three observing blocks. To
minimise flux variations among the different slicers and channels of the instrument in the
combined datacube, the telescope pointing was slightly offset, and the instrument's position
angle on the sky was rotated by $90\degr$ between each exposure. During each night, a
spectrum of the standard star HD\,49798 was recorded, also under photometric conditions.

The data reduction was performed using the MUSE pipeline v2.2 \citep{Weilbacher20} within
the ESO Recipe EXecution tool (EsoRex) environment \citep{ESOREX}.
Wavelengths in the observer's frame were converted to vacuum and corrected to
the solar-system barycentre. The standard-star frames were processed to derive
flux calibration solutions, and the sky background was measured in each science
exposure using regions of the field of view free of light contamination. Flux calibration
and sky subtraction were applied to the science frames during post-processing, and the
reduced pixel tables from each exposure were combined to produce a stacked datacube. The
absolute spectro-photometric accuracy is approximately 0.05 mag.

Finally, the residuals from sky emission lines were removed using the Zurich Atmosphere Purge
(ZAP) algorithm \citep{Soto16}, and the MUSE data product was aligned to match the astrometry of
the \hst\ images.
The astrometric registration was performed by identifying compact sources common to both the
MUSE white-light image and the \hst\ F814W image. A small rigid 2D shift ($\lesssim 0\farcs 2$)
was applied to the MUSE datacube World Coordinate System, yielding a relative astrometric
accuracy between MUSE and \hst\ of better than $0\farcs 1$.

\section{Image plane and lensing model}

\subsection{Galaxy census}\label{sec:census}

We conducted a blind survey for galaxies within the MUSE field of view. Preliminary source identifications
were performed on the deep 1.4-$\mu$m (F140W) \hst\ image using the Source Extractor and Photometry
software \citep{Bertin96,Barbary18}. We extracted 1D spectra from the MUSE datacube for each identified source
using \pymuse\ \citep{Pessa20}. Due to field crowding, particularly the presence of a foreground
cluster cD elliptical and a Galactic star to the southeast of it, we estimated the diffuse background light,
using concentric elliptical apertures around each source. The light was interpolated to the object inner region
and subtracted from the flux in each object spaxel before extracting their spectra. To produce a
combined 1D spectrum, we weighted each spaxel by their total integrated flux over the full MUSE wavelength
range, using the white-weighted mean option of \pymuse.

We also searched for \oii\,$\lambda\lambda$3727,3729 emission lines across the field at approximately the
redshift of the \mgii\ absorption lines detected in the \megasaura\ spectrum. This was done by constructing a MUSE
narrow-band image (70~\AA\ wide, corresponding to $\approx 2600$~km~s$^{-1}$ at $z=1.17$) centred on the
redshifted \oii\ doublet. We detected six \oii-emitting galaxies at $z\simeq 1.17$, two of which show weak or
no continuum emission. All of these galaxies have counterparts in the \hst\ optical images, indicating that
none were missed in the earlier source identification. We label them G1, G2, G3, G4, G5, and G6, in order
of increasing projected distance from G1. As shown in Fig.~\ref{fig:field}, these galaxies are roughly aligned on
the sky, suggesting the presence of a large-scale filamentary structure. G1, the largest and most massive
galaxy in the group (see Sect.~\ref{sec:masses}), lies near the centre of the field. Its closest
companion, G2, is located approximately $10\arcsec$ to the northeast of G1. The remaining four galaxies
are distributed on either side of the G1-G2 pair.

All sources identified at $z\gtrsim 1.17$, relevant to the analysis of gaseous absorption at $z\simeq 1.17$,
are listed in Table~\ref{tab:gal}. Object classifications and, where available, redshifts were determined using
the template-matching algorithm MARZ \citep{Hinton16}. In addition to the gravitational arc (hereafter
referred to as "the Arc") and the two counter-images (Ci1 and Ci2) of the $z\simeq 2.39$ lensed galaxy \sgas, the
field contains a quasar at $z=2.0966$, located approximately $5\arcsec$ north of the Arc, and five additional
galaxies (labelled with the prefix 'gal'), primarily at $z\simeq 1.43$.

\subsection{Lensing model and connection between G1 and G2}\label{sec:lens}

\begin{figure}
\begin{minipage}{1.\columnwidth}
\begin{center}
\includegraphics[width=0.9\columnwidth]{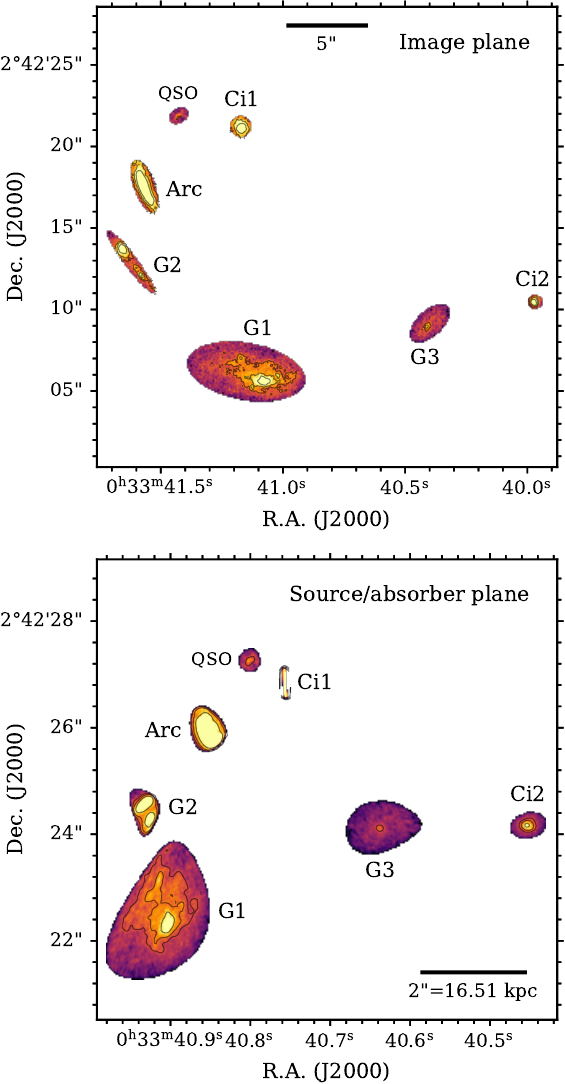}
\end{center}
\end{minipage}
\caption{Brightest sources at $z>1.16$ near the centre of the foreground galaxy cluster observed in \hst\ F140W imaging.
Black contours trace the flux distribution in the image.
The {\it upper panel} shows the image plane, while the {\it lower panel} displays the delensed absorber plane at $z=1.17$.
The elliptical aperture around G1 includes its eastern region, which is stretched by gravitational lensing.}
\label{fig:delens}
\end{figure}

G1 is a spiral galaxy, whose image is distorted by gravitational lensing. This effect, caused by the massive foreground
galaxy cluster, stretches the images and amplifies their luminosities. To recover an accurate representation of G1 and
other sources at the absorber's redshift, we developed a lens model for \sgas\ using the
\lenstool\ software \citep{Jullo07}, constrained by the \hst\ imaging.
The model, described in \citet{Fischer19} and available upon request, allows us to reconstruct the true positions of
the sources in the absorber plane by ray tracing through the lens equation. The relative accuracy of the
reconstructed positions varies across the field due to the gradient in magnification, resulting in an uncertainty
of 10\% in the calculated impact parameters. This level of uncertainty is typical of \arctomo\
studies \citep[e.g.,][]{Tejos21}. The magnification map at the absorber redshift of $z=1.17$ is shown in
Fig.~\ref{fig:magni}.

Applying the deflection matrices to the \hst\ F140W image recovers the absorber-plane geometry shown in
Fig.~\ref{fig:delens}. This reveals that G1 and G2 are closely aligned in projection, with a separation of less
than 15~kpc, and appear less elongated than in the image plane. The high-resolution \hst\ image shows
a double-peaked light profile in G2, which is more compact than the extended, diffuse emission from G1.
Due to the dazzling glare of the Galactic star partially obscuring this region,
the presence of a bridge of matter connecting the two galaxies is not immediately apparent (see Sect.~\ref{sec:maps}).
Nevertheless, G2 is likely an interacting companion of G1. The approximate alignment of their delensed images, along with
the presence of a bar in G1's core, strongly supports this interpretation (see Fig.~\ref{fig:delens}).

\section{Emission properties of group galaxies}\label{sec:emi}

To characterise the group at $z\simeq 1.17$, we first combined the \hst\ and MUSE datasets to constrain the
stellar masses, star formation rates, and disc properties of the individual galaxies.

\subsection{Stellar masses}\label{sec:masses}

We determined the stellar mass of each galaxy from the combination of \hst\ photometry and MUSE
continuum spectra. For this, we employed the Bayesian Analysis of Galaxies for Physical Inference
and Parameter EStimation software, \bagpipes\ \citep{Carnall18,Carnall19}. \hst\ magnitudes were
measured in the image plane and corrected for Galactic extinction \citep{Schlafly11}.
The stellar population models used in \bagpipes\ were the 2016 versions of
the \citet{Bruzual03} models, incorporating the stellar initial mass function (IMF)
of \citet{Kroupa01}. Star-formation histories were modelled as double-power-laws
following \citet{Behroozi13}. After correcting for a median magnification factor of $\mu=5.6$ (see
Table~\ref{tab:sfr}),\footnote{Since our SED fitting uses integrated photometry and
spectra, the derived quantities represent magnification-weighted averages. Differential lensing
could introduce systematic uncertainties in stellar mass and SFR estimates when the
magnification varies across a galaxy's image (see Fig. B.1). For G1, the gradients are typically
of order 20-30\% across its extent. We estimate this introduces a small systematic uncertainty of
$\sim 0.1$~dex in stellar mass, which is comparable to the statistical uncertainties reported in
Table~\ref{tab:sfr}. The impact on G1's SFR estimate is similar, as both stellar mass and
\oii-based SFRs are corrected using the same median magnification factor. Such systematic
effects are much smaller for the less extended G2 to G6 galaxies.}
we inferred a stellar mass of $\log (M_\star/M_\odot)=11.01\pm 0.08$ for G1. Based on the
stellar-to-halo mass relation of \citet{Moster10}, this corresponds to a
virial radius of $R_{200}=271\pm 33$~kpc. We applied the same methodology to the other
identified group members. Our results are summarised in Table~\ref{tab:sfr}.

From this, we conclude that G1 and G3 are the most massive galaxies in the group and that the virial radii of
all galaxies, except G6, overlap, indicating a bound structure. It remains to be assessed whether each galaxy in this
structure maintains its own CGM or shares a common halo.

\subsection{Star-formation rates}\label{sec:sfr}

\begin{table*}
\caption{Galaxy masses and star-formation rates.}\label{tab:sfr}
\centering
\begin{tabular}{ccccccccc}
\hline\hline
& $\log (M_\star/M_\odot)$\tablefootmark{a} & $\log (M_\mathrm{h}/M_\odot)$ & $R_\mathrm{vir}$ &
{\it SFR}$_\mathrm{SED}$\tablefootmark{a} & $A_\mathrm{V}$\tablefootmark{a,b} & $\mu$\tablefootmark{c} & $f_\mathrm{[OII]}$\tablefootmark{d,e} & {\it SFR}$_\mathrm{[OII]}$\tablefootmark{f} \\
& & & [kpc] &
[$M_\odot$\,yr$^{-1}$] & (mag) & & & [$M_\odot$\,yr$^{-1}$] \\
\hline
G1 & $11.01\pm 0.08$ & $12.93\pm 0.16$ & $271\pm 33$ &
$63.7\pm 0.1$ & 1.13 &  5.6 &  $842\pm 19$  & $70.0\pm 1.6$ \\
G2 & $10.14\pm 0.06$ & $11.96\pm 0.03$ & $129\pm 3$  &
$4.9\pm 0.1$ & 0.38 & 10.5 & $66.9\pm 0.6$ &  $4.7\pm 0.1$ \\
G3 & $10.56\pm 0.12$ & $12.26\pm 0.12$ & $162\pm 14$ &
$13.4\pm 0.1$ & 1.42 &  3.7 &  $216\pm 8$   & $17.4\pm 0.7$ \\
G4 &  $9.56\pm 0.18$ & $11.69\pm 0.08$ & $105\pm 6$  &
$1.4\pm 0.6$ & 0.68 &  2.1 & $32.4\pm 2.1$ &  $1.8\pm 0.2$ \\
G5 &  $9.44\pm 0.16$ & $11.64\pm 0.07$ & $101\pm 5$  &
$17.7\pm 0.2$ & 1.71 &  2.1 &  $386\pm 7$   & $19.7\pm 0.4$ \\
G6 & $10.22\pm 0.22$ & $12.00\pm 0.13$ & $134\pm 13$ &
$1.3\pm 1.4$ & 0.34 &  1.5 & $23.9\pm 1.3$ &  $1.7\pm 0.1$ \\
\hline
\end{tabular}
\tablefoot{
\tablefoottext{a}{Derived from SED fitting;}
\tablefoottext{b}{Internal extinction in V band;}
\tablefoottext{c}{Median magnification factor;}
\tablefoottext{d}{\oii\ flux corrected for gravitational magnification, Galactic extinction, and internal extinction;}
\tablefoottext{e}{In units of $10^{-18}$~erg\,s$^{-1}$\,cm$^{-2}$;}
\tablefoottext{f}{Star-formation rate based on \oii\ emission.}}
\end{table*}

To derive star-formation rates (SFRs), we used the Python Spectroscopic Analysis and Plotting package, \pyspeckit\
\citep{Ginsburg22}. We fitted the \oii\,$\lambda\lambda$3727,3729 emission line doublet in the previously extracted
MUSE spectra of G1 to G6 using double Gaussian profiles on top of normalised continua (see Fig.~\ref{fig:spec1}).
The measured fluxes were corrected for gravitational magnification, Galactic extinction at the observed wavelengths
(using the reddening maps from \citealt{Schlafly11}), and internal galaxy extinction at rest-frame wavelengths assuming
the \citet{Calzetti00} extinction law for starbursts and the internal extinction in $V$ band, $A_\mathrm{V}$, derived from
\bagpipes\ (see Sect.~\ref{sec:masses}). The total \oii\ luminosities were converted to SFRs following
\citet{Davies16}, but assuming the IMF of \citet{Kroupa01}, using the following relation:
\begin{equation}\label{eq:sfr}
\mathrm{SFR}_\mathrm{[OII]}~(\mathrm{M}_\odot\,\mathrm{yr}^{-1}) = 1.1 \times 10^{-41} L_\mathrm{[OII]}~(\mathrm{erg\,s}^{-1}).
\end{equation}
When converting the observed luminosities to intrinsic values, we applied an offset
of $0.1$~dex towards lower metallicities at $z\approx 1$ relative to the local mass-metallicity relation
\citep[e.g.,][]{Lara-Lopez13,Sanders21}. Our results are summarised in Table~\ref{tab:sfr}. For G1, we derived
an unobscured SFR of \sfroii~$M_\odot$~yr$^{-1}$. For comparison, G1's integrated SFR based on SED fitting
(see Sect.~\ref{sec:masses}) is \sfrsed~$M_\odot$~yr$^{-1}$. These differences are expected, as instantaneous SFR traces
recent star formation over $\sim 10$~Myr, while SED fitting averages over longer
timescales \citep[$\gtrsim 100$~Myr; e.g.,][]{Kennicutt12}. For the other five galaxies, the two methods yield
results differing by up to 30\%, typically indicating enhanced recent star formation activity.

Most galaxies in the group follow the main sequence of star formation, which links the star-formation rate to the
stellar mass, at intermediate redshifts \citep[e.g.,][]{Bauer05}. However, G1 is positioned at the high end of the
SFR distribution for its relatively large stellar mass. Its specific SFR is $0.69$~Gyr$^{-1}$ -- a factor of 20 higher
than the galaxy main sequence at $z=1.2$ \citep{Bauer05} and three times higher than the average starburst galaxy at
that redshift \citep{Atek14,Pearson18}. Given G1's unique configuration within the group, it appears to be experiencing a
starburst phase driven by a small number of massive star-forming regions. These clumps are likely the result of violent
disc instabilities triggered by interactions \citep[e.g.,][]{Elmegreen09,Puschnig23}.

\subsection{Morpho-kinematics of [OII] emission}\label{sec:galpak}

\begin{table*}
\caption{Morpho-kinematic analysis of [\ion{O}{ii}] emission.}\label{tab:models_Gs}
\centering
\begin{tabular}{clcccccc}
\hline\hline
& Coords. R.A. \& Dec.\tablefootmark{a} & PA$_\mathrm{gas}$\tablefootmark{a,b} & $i_\mathrm{gas}$\tablefootmark{a} & $r_\frac{1}{2}$\tablefootmark{a,c} & $z_\mathrm{sys}$ & $v_\mathrm{rot,gas}$\tablefootmark{a,d} & $\sigma_{v\mathrm{,gas}}$\tablefootmark{a,d} \\
& (J2000) & [\degr] & [\degr] & [kpc] & & [\kms] & [\kms] \\
\hline
G1 & $0^{\rm h} 33^{\rm m} 40\fs 90$ $+2\degr 42\arcmin 22\farcs2$\tablefootmark{e}
                                                                   & $149.1\pm 0.5$ & $49.5\pm 4.7$ & $7.4\pm 0.01$ & $1.16666$ & $276.8\pm 2.1$  & $26.2\pm 2.5$ \\
G2 & $0^{\rm h} 33^{\rm m} 40\fs 92$ $+2\degr 42\arcmin 24\farcs3$\tablefootmark{e}
                                                                   & $43.8\pm 0.7$
                                                                   & $75.8\pm 0.6$ & $1.6\pm 0.02$ & $1.16745$ & $142.2\pm 2.5$  & $74.2\pm 1.0$ \\
G3 & $0^{\rm h} 33^{\rm m} 40\fs 63$ $+2\degr 42\arcmin 24\farcs1$\tablefootmark{e} & $128.1 \pm 2.7$ & $42.0\pm 0.0$\tablefootmark{e}  & $2.2\pm 0.09$ & $1.16484$ & $169.9\pm 5.1$ & $\vec{<8.4}$\tablefootmark{f} \\
\hline
\end{tabular}
\tablefoot{
\tablefoottext{a}{Delensed absorber plane;}
\tablefoottext{b}{Position angle of the approaching semi-major axis (east of north);}
\tablefoottext{c}{Half-light radius;}
\tablefoottext{d}{Assuming a rotation curve $v(r)=v_\mathrm{rot,gas}\frac{2}{\pi}\arctan(r/r_\mathrm{t})$;}
\tablefoottext{e}{Fixed parameter;}
\tablefoottext{f}{Upper limit: the \oii\ emission is unresolved at MUSE spectral resolution.}}
\end{table*}

The ubiquity of \oii\ emission among the galaxies in this group provides a
valuable opportunity to study the kinematics of ionised gas directly associated with
star formation. Unlike the other group members, G1, G2, and G3 are spatially resolved and
extended enough to allow for the extraction of meaningful kinematic data. To model their
dynamics, we employed the Galaxy Parameters and Kinematics from 3-Dimensional data
software, \galpak\ \citep{Bouche15}, fitting the MUSE data using single rotating disc models
characterised by an exponential light profile, an arctangent rotation curve, and a
Gaussian thickness profile.
Following the methodology outlined in \citet{Tejos21}, we used the delensed \oii\
emission-line MUSE datacube and an effective delensed PSF as input to \galpak. The delensed
cube was constructed at the absorber-plane redshift by applying the deflection matrices to
the vertices of each spaxel, within a sub-cube of the original data, centred on each galaxy and
with the spectral continuum subtracted. The spaxels in the delensed sub-cube were then
resampled onto a regular $0.1\times 0.1$~arcsec$^2$ grid, consistent with the typical size of
the delensed spaxels in the absorber plane.
This regridding introduces correlations between adjacent spaxels in those regions or
directions where delensing is negligible. This increases the noise correlation length but
does not bias the derived kinematic parameters, as \galpak\ samples the model at the native
spaxel positions. The spaxel correlation therefore primarily affects the goodness-of-fit
statistics rather than the best-fit parameters.
Care was taken to define extraction regions around each galaxy, avoiding spaxels subject
to strong tangential stretching, especially in the area between G1 and G2.
The spatially variable PSF in the absorber plane was accounted for by
providing \galpak\ with the local effective PSF at each galaxy's position. We verified
that using a constant PSF changes the derived rotation velocities by less than 5~\kms, well
within our reported uncertainties.

\galpak\ was run to convergence for up to 10,000 iterations, yielding satisfactory kinematic
solutions for all three galaxies. The modelling of G2, however, was subject to
larger uncertainties owing to the limited number of available MUSE spaxels. To improve convergence,
we fixed its
central coordinates to the mid-point of the two brighter peaks in the delensed \hst\ image.
Note that, for illustrative purposes, the image of G2 in Fig.~\ref{fig:delens} is truncated;
the galaxy extends slightly further to the south-west. For G1 and G3, we also fixed
their central coordinates based on the delensed \hst\ image, while allowing all other parameters
to vary, except the inclination of G3, which was determined from its major-to-minor axis ratio.
However, these adjustments had only minor effects on the results.
From these analyses, we derived the key disc parameters of each galaxy, including the
systemic redshift ($z_\mathrm{sys}$), maximum rotational velocity ($v_\mathrm{rot,gas}$),
and gas velocity dispersion ($\sigma _{v,\mathrm{gas}}$), as summarised in
Table~\ref{tab:models_Gs}. The systemic redshifts obtained with \galpak\ are more accurate
than those derived from simple Gaussian fits to the integrated \oii\ emission lines
(see Sect.~\ref{sec:sfr}) and are adopted hereafter. The inferred rotational velocities are
consistent with expectations based on the galaxies' masses. G1 and G3 exhibit well-ordered,
star-forming discs, with large $v_\mathrm{rot}/\sigma_v$ ratios,
whereas G2 shows signs of disturbed rotation, likely reflecting G1's dynamical influence.

\section{Absorption lines at $z\simeq 1.17$}\label{sec:abs}

\begin{table*}
\caption{Absorption lines at $z\simeq 1.17$: properties from MUSE integrated spectra.}\label{tab:abs}
\centering
\begin{tabular}{lllllllll}
\hline\hline
Object & Beam type & $d_{\rm proj}$\tablefootmark{a} & $\Delta v$\tablefootmark{b} & $\sigma _v$\tablefootmark{c}   &
$W^{2796}_\mathrm{r}$                                           &
$\frac{W_\mathrm{r}(\lambda 2803)}{W_\mathrm{r}(\lambda 2796)}$ &
$\frac{W_\mathrm{r}(\lambda 2600)}{W_\mathrm{r}(\lambda 2796)}$ &
$\frac{W_\mathrm{r}(\lambda 2852)}{W_\mathrm{r}(\lambda 2796)}$ \\
       &           & [kpc]                           & [km~s$^{-1}$]               & [km~s$^{-1}$]                   &
[\AA]                                         &
                                              &
                                              &
                                              \\
\hline
G1       & DtB/extended         &  15.1 & $-21\pm 10$ & $151\pm 9$  & $3.61\pm 0.32$ & $1.11\pm 0.14$ & $0.99\pm 0.16$ & $0.28\pm 0.07$ \\
G2       & DtB/extended         &   4.9 & $+20\pm 11$ & $156\pm 9$  & $4.16\pm 0.40$ & $1.02\pm 0.11$ & $0.79\pm 0.15$ & $0.63\pm 0.10$ \\
G3       & DtB/extended         &  40.1 & $-30\pm 47$ & $245\pm 36$ & $4.63\pm 1.00$ & $0.66\pm 0.23$ & $0.70\pm 0.27$ & $0.38\pm 0.16$ \\
G5       & DtB/extended         &  64.6 &      ...    &     ...     & $<2.28$\tablefootmark{d} & ...  &      ...          &      ...       \\
\hline
Arc      & extended             &  20.8 & $+61\pm 2$  & $150\pm 2$  & $4.09\pm 0.05$ & $0.99\pm 0.02$ & $0.78\pm 0.02$ & $0.35\pm 0.02$ \\
Ci1      & extended             &  34.0 & $+16\pm 7$  & $134\pm 6$  & $3.70\pm 0.15$ & $0.98\pm 0.07$ & $0.75\pm 0.09$ & $0.36\pm 0.05$ \\
Ci2      & extended             &  62.6 & $-33\pm 11$ & $132\pm 11$ & $3.54\pm 0.40$ & $0.87\pm 0.14$ & $0.46\pm 0.13$ & $0.22\pm 0.08$ \\
\hline
QSO      & point source         &  32.8 & $+34\pm 18$ & $71\pm 15$  & $2.30\pm 0.39$ & $1.00\pm 0.24$ & $0.99\pm 0.47$ & $0.55\pm 0.24$ \\
gal2     & extended             &  95.2 &     ...     &    ...      &    $<1.30$\tablefootmark{d} & ...  &      ...       &      ...       \\
gal3     & $\sim $point source  & 163.5 &     ...     &    ...      & $\le 1.44$\tablefootmark{e} & ...  &      ...       &      ...       \\
\hline
\end{tabular}
\tablefoot{
\tablefoottext{a}{Projected distance from the group's barycentre (absorber plane);}
\tablefoottext{b}{Relative to the mass-weighted mean redshift of the group ($z_{\rm sys}=1.16663$);}
\tablefoottext{c}{Deconvolved from the instrumental Line Spread
Function ($\sigma _\mathrm{inst}=1.083$~\AA\ at $\lambda_\mathrm{obs}=6060$~\AA);}
\tablefoottext{d}{$2\sigma$ upper limit;}
\tablefoottext{e}{Blend.}}
\end{table*}

With all relevant galaxy parameters established, we now turn to the central aim of this study:
investigating the properties of cool-gas absorption within the group.

\subsection{Down-the-barrel absorption}\label{sec:intervening}

We find that half of the galaxies in the $z\simeq 1.17$ group display down-the-barrel
(DtB) absorption, meaning that absorption is detected at approximately the same redshift
as the galaxies themselves, superimposed on their continuum emission spectra
(see Table~\ref{tab:gal}). This absorption is most clearly observed towards G1 and G2, but
is also present towards G3. Such features probe the CGM and/or IGrM located in front of each
galaxy, as well as their interstellar media (ISM).

The DtB \mgii\ absorption is blueshifted with respect to the galaxies' systemic redshifts,
with velocity shifts of $-40\pm 10$ and $-94\pm 11$~\kms\ for G1 and G2, respectively (see
Table~\ref{tab:abs} and Fig.~\ref{fig:spec2}). This indicates the presence of outflowing gas,
possibly driven by star formation. However, the bulk of the absorption in front of G1,
where the signal is better resolved than for G2, is not aligned with G1's minor axis.
Instead, it is predominantly oriented towards G2 (see Fig.~\ref{fig:velmaps_absplane}). This
geometry suggests that gas is being stripped as a result of the interaction between the two
galaxies, causing each to lose part of their gaseous reservoirs. Although the region between
G1 and G2 is partly obscured by the glare of the Galactic star, there are also indications of
a bridge of material traced by \oii\ emission connecting the two
galaxies (Figs.~\ref{fig:velmaps_absplane}). We interpret this as the region where material is
transferred into their shared CGM. Shock-heated gas within tidal features is typically observed
in interacting systems \citep[e.g.,][]{Smith10,Joshi19}. Therefore, while star-formation-driven
winds (particularly from G1, given its elevated SFR) may contribute to the outflow,
tidal stripping is likely to play a significant role in removing gas from these
galaxies \citep[see][]{Sparre22}.

For G3, the DtB \mgii\ absorption exhibits a complex profile (see Fig.~\ref{fig:spec2}), with
an average velocity shift of $+208\pm 47$~\kms\ relative to the galaxy's systemic redshift.
This positive offset cannot be attributed to outflowing gas. Instead, it shows that a
significant fraction of the absorption originates at approximately the velocity
of G1 ($+258$~\kms\ relative to G3), i.e., within the IGrM, as G3 moves towards the observer.

\subsection{Intervening intragroup absorption}

Intervening \mgii\ absorption towards sources at higher redshifts than the group
itself pervades the field. Such absorption is detected towards all extragalactic background
sources at $z>1.17$, spanning projected distances from G1 of up to 70~kpc in the absorber
plane (see Fig.~\ref{fig:spec2}). Note that these sources preferentially probe regions along
G1's major axis.

Intervening \mgii\ absorption towards the Arc and its two counter-images, Ci1 and Ci2, are
reminiscent of ultra-strong \mgii\
absorbers \citep[e.g.,][]{Nestor07,Nestor11,Rubin10,Gauthier13,Nielsen22,Guha24} observed
towards quasars, with rest-frame equivalent widths of the \mgii\,$\lambda$2796 line in excess
of 3~\AA. This is surprising since a wide range of impact parameters from G1 are probed, i.e.,
between 30 and 60~kpc, in the delensed absorber plane. For sources farther than 70~kpc,
only upper limits are derived. Line equivalent widths and ratios involving the
\mgii\,$\lambda$2796 line, derived from MUSE integrated spectra, are given
in Table~\ref{tab:abs}. The equivalent widths are subject to uncertainties related to
the subtraction of diffuse background light around each source, which was performed
during spectrum extraction (see Sect.~\ref{sec:census}). This introduces an additional
uncertainty of up to 20\% on the reported values.

In the spectrum of the Arc, we confirm the detection of \mgii\,$\lambda\lambda$2796,2803 and
identified additional, prominent absorption lines at $z\simeq 1.1671$, i.e., \mgi\,$\lambda$2852,
\feii\,$\lambda\lambda$2249,2260,2344,2374,2382,2586,2600, and
\caii\,$\lambda\lambda$3934,3969. Neither \mnii\,$\lambda$2576 nor \nai\,$\lambda$5891
are detected. The strengths of the detected lines are reminiscent of what is observed
in high-redshift quasar-DLAs.
\citet{Berg25} used \mgii\ metrics coupled with modelling to demonstrate that
this system is indeed likely a DLA, with $\log N($\hi$)\simeq 20.6$~cm$^{-2}$, and
a covering fraction on top of the Arc of approximately 70\%.

\subsection{Spatially extended versus pencil-beam sightlines}\label{sec:compa_ext_vs_ps}

\begin{figure*}
\begin{center}
\includegraphics[width=0.9\columnwidth]{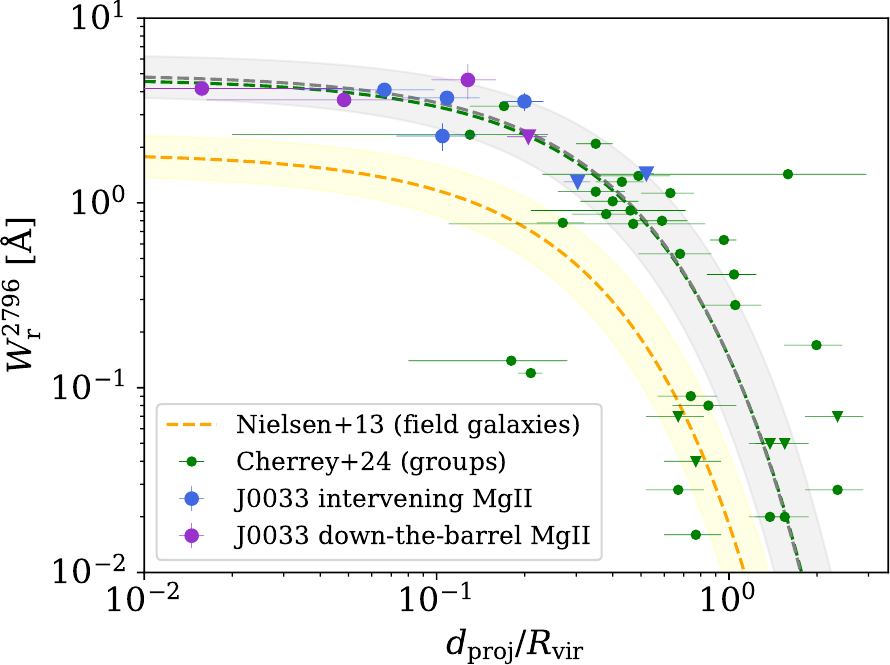}\hspace{0.1\columnwidth}
\includegraphics[width=0.9\columnwidth]{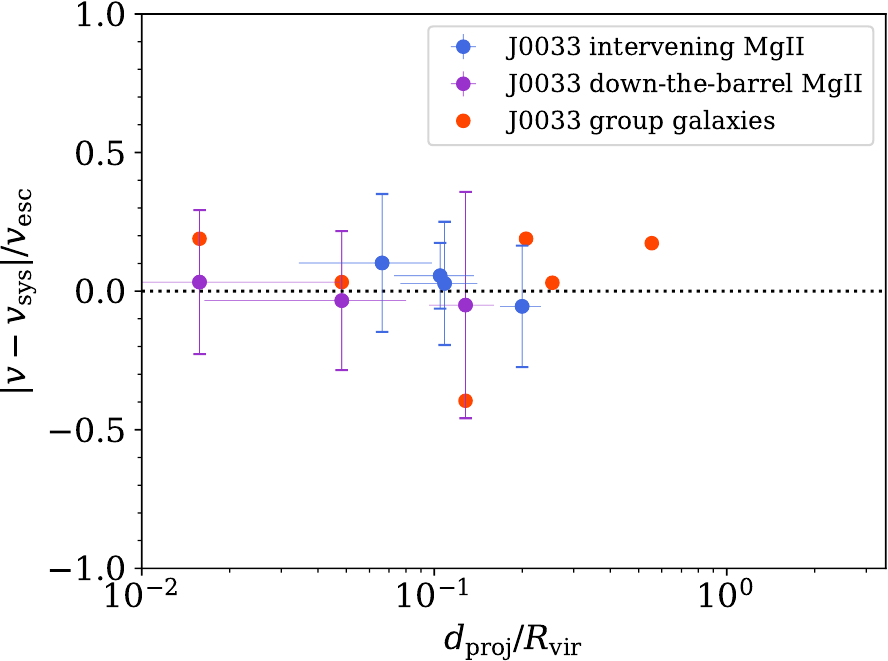}
\end{center}
\vspace{-0.05\columnwidth}
\caption{
{\it Left panel:} Rest-frame equivalent width of \mgii\,$\lambda$2796 as a function of the
projected distance from the group's barycentre, normalised by the group's virial radius.
Our measurements are compared to the galaxy group sample from \citet{Cherrey24}. The overall
best-fit relation for {\it intervening} \mgii\ absorption is displayed as a grey dashed line along
with its $1\sigma$ confidence region. The original relation from \citet{Cherrey24} is shown
in green.
{\it Right panel:} Velocity span of the \mgii\ profiles, $\Delta v\pm \sigma_v$, normalised by the
group's escape velocity, $v_{\rm esc}=601$~\kms, as a function of the normalised projected
distance from the group's barycentre. The velocities of the individual galaxies are shown in
red. The reference velocity, $v_{\rm sys}$, is the mass-weighted mean redshift of the
group ($z_{\rm sys}=1.16663$).}
\label{fig:plot_boRvir}
\end{figure*}

\mgii\ absorption is detected at the $5\sigma$ confidence level in the spectrum of a quasar at
$z\simeq 2.10$, located in projection near both the Arc and Ci1. The quasar sightline
serves as a pencil beam probe, while the Arc and Ci1 correspond to spatially
resolved, extended background sources. To our knowledge, this is the first time that both
beam types can be compared at projected separations as small as $5-10$~kpc, thereby probing
the same physical environment.

The \mgii\,$\lambda\lambda$2796,2803 doublet in the quasar spectrum is saturated, as indicated
by its equivalent-width ratio (see Table~\ref{tab:abs}), consistent with what is observed in
the spectra of the Arc and Ci1. After subtracting the diffuse background light around
each source, the absorption troughs reach zero flux within the uncertainties of
the subtraction process ($\sim 10$\%, or $\sim 20$\% for the fainter quasar;
see Fig.~\ref{fig:spec2}). This suggests no evidence for partial covering of the
background sources by the \mgii-absorbing gas.

Both the equivalent widths and velocity widths of the \mgii\ lines are larger towards the
extended sources (the Arc, Ci1, and Ci2) than along the quasar sightline. This comparison
is based on a single quasar spectrum, whose limited S/N precludes a detailed analysis
of line-profile shapes. Nevertheless, because \arctomo\ spectra are spatially averaged analogues
of pencil-beam data, they are expected to provide a more global view of the intervening
medium, with reduced sightline-to-sightline variations compared to quasars, particularly in
the presence of clumpy gas distributions.

\section{Results}
\label{sec:results}

\subsection{Radial distribution of cool gas in groups}

\citet{Cherrey24} investigated intervening \mgii\ absorption along quasar sightlines in the
vicinity of 26 galaxy groups, each containing more than five members, spanning
redshifts $0.3<z<1.5$ and halo masses $10.7<\log(M_\mathrm{h}/M_\sun)<13.7$. Out of 120
absorption systems, 21 were associated with such groups. They found that the typical impact
parameter from the nearest galaxy, at which the \mgii\ covering fraction exceeds 50 per cent,
is about three times larger than that for field galaxies. This indicates the presence of an
IGrM more extended than the CGM of individual group members \citep[see
also][]{Bordoloi11,Bielby17}.

Our tomographic observations of cool ($\sim 10^4$~K) gas in the field of \sgas\ provide a
unique opportunity to probe the IGrM from within, at smaller impact parameters than
those accessible through quasar sightlines. As shown in the left panel of
Fig.~\ref{fig:plot_boRvir}, the \arctomo\ measurements, obtained at projected distances of
less than one-fifth of the group's virial radius, follow the trend reported by
\citet{Cherrey24}. A joint fit to the {\it intervening} \mgii\ data from both the
\arctomo\ and Cherrey et al. samples yields:
\begin{equation}\label{eq:newcherrey}
\log W_\mathrm{r}^{2796}~[\AA] = (-1.53\pm 0.28)\times d_\mathrm{proj}/R_\mathrm{vir} + 0.70\pm 0.22.
\end{equation}
The low scatter observed across the inner group halo suggests a well-mixed IGrM,
although this may also reflect the large \arctomo\ beams. The DtB \mgii\ absorption
detected in front of G1 and G2 falls along this relation, indicating that these
sightlines probe velocity dispersions consistent with those of intervening systems.
At larger projected distances, beyond one-fifth of the virial radius, no \mgii\
absorption is detected from \arctomo, consistent with both the declining trend and
our relatively loose upper limits.

Compared to field galaxies \citep{Nielsen13}, galaxy groups exhibit
enhanced \mgii\ absorption, implying that substantial amounts of cool gas have
been displaced from galaxies into the IGrM through feedback processes.
In the right panel of Fig.~\ref{fig:plot_boRvir}, we show the velocity ranges of gas clouds
along each sightline, measured relative to the group's mass-weighted systemic velocity.
All measured velocities lie well below the group's escape velocity, indicating that the cool
gas is strongly gravitationally bound to the group.

\subsection{Chemical enrichment}

\begin{figure*}
\begin{center}
\includegraphics[width=0.95\hsize]{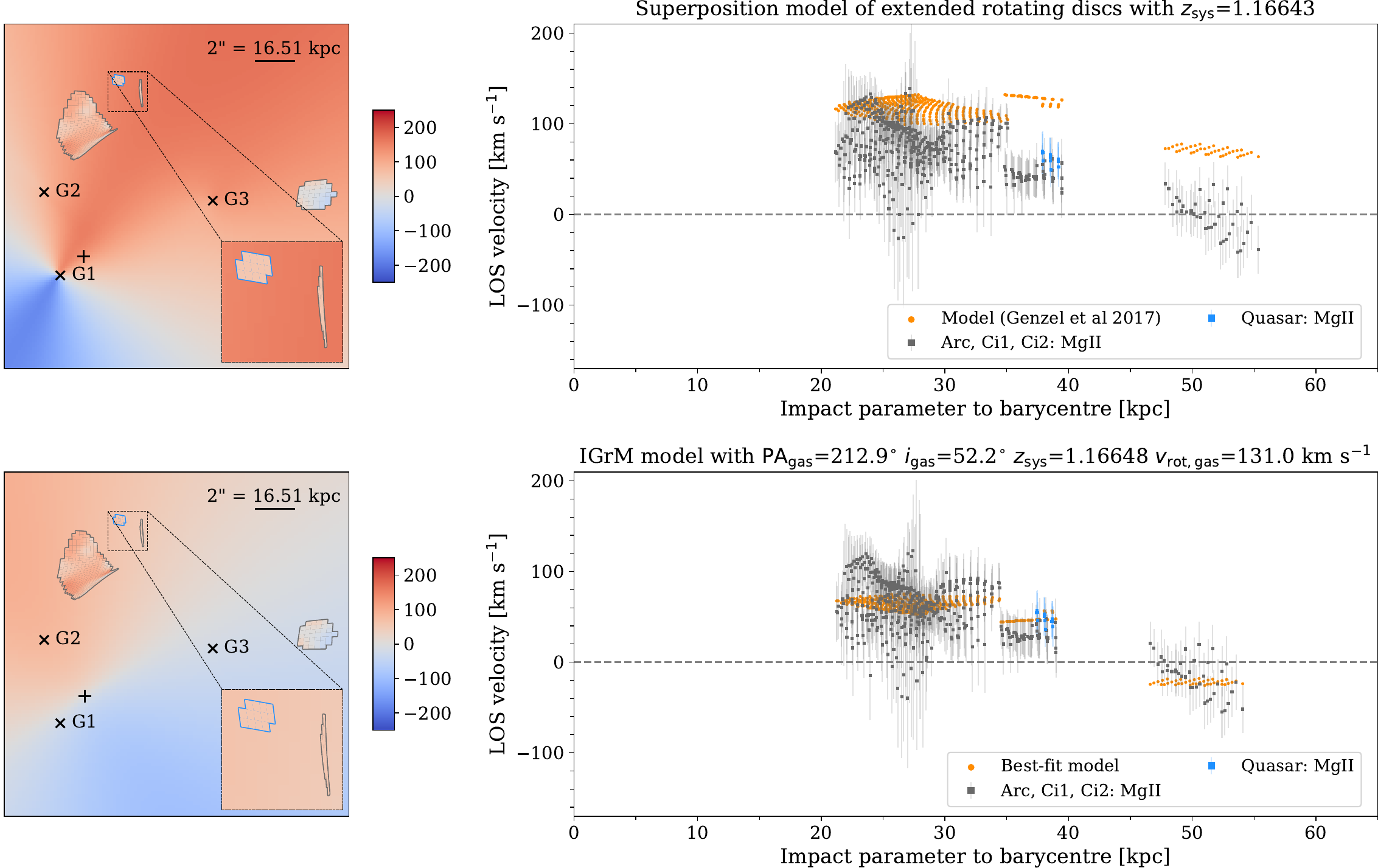}
\end{center}
\vspace{-0.04\columnwidth}
\caption{Superposition model of extended rotating discs ({\it upper panels}) and rotating IGrM
model ({\it lower panels}). {\it Left panels:} Modelled velocity maps with the
observed intervening \mgii\ mean velocities overlaid for each of the four delensed sources. The
inset shows the zoom-in on a region encompassing Ci1 and the quasar. {\it Right
panels:} Model-predicted velocities (orange) at the positions of the Arc, Ci1, Ci2, and
the quasar, as a function of impact parameter relative to the barycentre, marked by a plus
sign in the left panels (superposition model: barycentre of G1 to G3; IGrM: G1 to G5). The
observed \mgii\ velocities are displayed in black for the Arc, Ci1, and Ci2, and in blue for
the quasar.}
\label{fig:m_model}
\end{figure*}

The chemical enrichment of the gas can be probed through the relative abundances of iron and
magnesium at different locations across the field. In the case of \sgas, most of the \feii\
and \mgii\ absorption lines are strongly saturated, implying that their equivalent
widths primarily trace the velocity dispersion of individual clouds along each sightline.
Across all background sources, the (Fe/Mg) ratio remains consistent within the
measurement uncertainties, except towards Ci2, where the \feii\ lines are weaker
(see Table~\ref{tab:abs}). In this region, the ratio suggests a lower iron abundance relative
to magnesium and/or a reduced metallicity.
Since iron is predominantly produced by Type Ia supernovae on longer timescales than the
alpha-elements like magnesium (produced in core-collapse supernovae), a lower Fe/Mg ratio
indicates younger or less chemically evolved gas \citep[e.g.,][]{Matteucci86}.
Overall, the IGrM in the field
of \sgas\ appears chemically homogeneous, with a possible decline in metallicity
towards the outer regions. This interpretation is consistent with the non-detection
of \mgii\ beyond a projected distance of 65~kpc from the group's barycentre.

\subsection{Gas kinematics}\label{sec:kinematics}

\begin{figure*}
\sidecaption
\includegraphics[width=0.67\hsize]{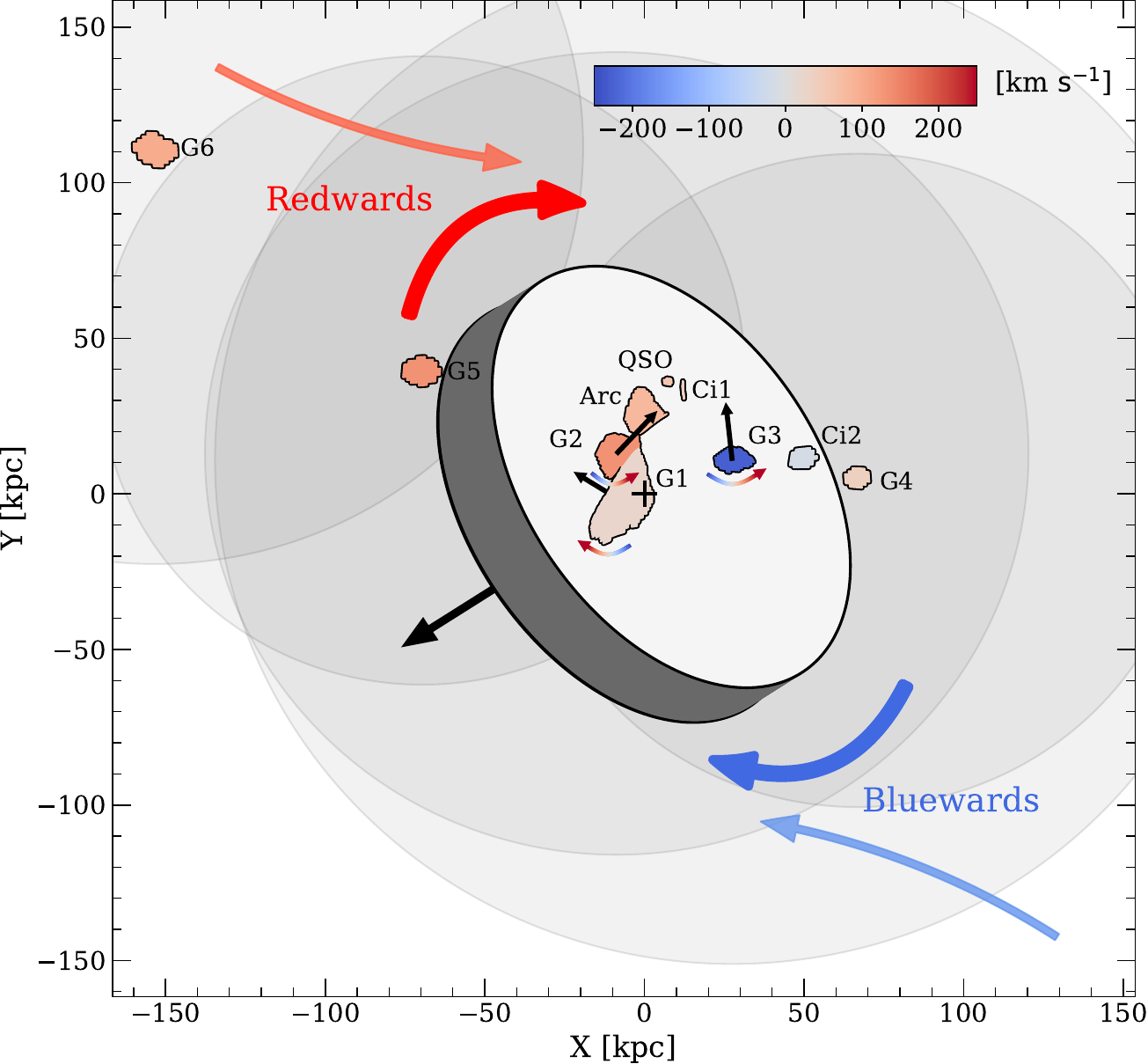}
\caption{Schematic of the orientation and kinematics of the intragroup medium in the compact galaxy group in
the field of \sgas. Each region is colour-coded based on the mean velocity offset of the
observed \mgii\ (or \oii) gas relative to the mass-weighted mean redshift of G1 to G5 ($z=1.16648$).
The axes (black arrows) and directions (blue and/or red arrows) of rotation of G1, G2, G3, and the
intragroup medium are indicated. The shaded, overlapping circles denote the extent of one virial
radius around each galaxy.}
\label{fig:schematic}
\end{figure*}

The detected intervening \mgii\ absorption lies predominantly along the projected major
axis of G1, making our non-down-the-barrel observations largely insensitive to
star-formation-driven outflows, which are expected to occur along the minor axis. To
investigate the origin of this absorption, we considered three scenarios: (1) extended
gaseous discs associated with G1, G2, or G3; (2) a superposition of such discs; and
(3) an IGrM halo shared by the group.

Using \galpak\ modelling (see Sect.~\ref{sec:galpak}), we extrapolated
the \oii\ emission-line kinematics of each galaxy
and compared the resulting velocity fields with the \mgii\ absorption. We adopted
realistic rotation curves from \citet{genzel20}, which account for the
increasing velocity dispersion in the outer disc regions. We find that single-disc models
fail to reproduce the mean absorption velocities with high confidence
(see Fig.~\ref{fig:g_model}), with reduced $\chi^2_\nu$ values of 17.7, 9.61, and 154
for G1, G2, and G3, respectively. This discrepancy is striking for G1, the most
massive galaxy in the group, which should dominate the absorption.
Similarly, G2's kinematics alone do not match the observations, despite its
close projected distance to the Arc. Nevertheless, G2's extrapolated velocity field
shows better agreement with the \mgii\ absorption in front of the Arc than G1's,
offering clues to the gas origin.
We then considered a superposition model by calculating the mass-weighted mean
velocities from the extended disc models of G1, G2, and G3 across the field. This
approach improves the match near the Arc ($\chi^2_\nu=4.43$) but fails at larger distances
(see Fig.~\ref{fig:m_model}, upper panels).\footnote{Note that the regions shown
in Fig.~\ref{fig:m_model} represent all spaxels where \mgii\ absorption is detected above our
significance threshold ($\mathrm{S/N}>2$ in the continuum). The Arc appears more extended than
in the lower panel of Fig.~\ref{fig:delens} because the spatial resolution of MUSE is lower than
that of \hst. Moreover, Ci1 appears narrow in both cases due to the geometry of the
delensing transformation, which compresses this image along one dimension (east-west) in
the absorber plane.} Overall, the \mgii\ gas exhibits kinematics
more similar to those of the least massive galaxies than to G1's. While this confirms
some level of association with individual galaxies, it is weaker than anticipated.
Indeed, the \mgii\ gas is confined to a narrow velocity range of $\sim\pm 50$~\kms\ around
the group's systemic redshift, much smaller than the stellar rotational velocities. Furthermore,
in the direction of Ci2, the gas counter-rotates relative to the ISM of both G1 and G2
(see Fig.~\ref{fig:g_model}), challenging the validity of the superposition model beyond a
first approximation.

\begin{table}
\caption{Orientation-kinematic analysis of \mgii\ absorption.}\label{tab:models_Mg}
\centering
\begin{tabular}{llllll}
\hline\hline
Rotation & PA$_\mathrm{gas}$\tablefootmark{a,b} & $i_\mathrm{gas}$\tablefootmark{a} & $z_\mathrm{sys}$\tablefootmark{c} & $v_\mathrm{rot,gas}$\tablefootmark{a,d} & $\chi^2_\nu$\tablefootmark{e} \\
centre & [\degr] & [\degr] & & [\kms] & \\
\hline
G1                & $231^{+1}_{-2}$ & $42^{+10}_{-13}$ & 1.16666 & $150^{+36}_{-13}$ & 0.711\\
G1+G2             & $238^{+1}_{-1}$ & $40^{+13}_{-12}$ & 1.16674 & $151^{+45}_{-17}$ & 0.713\\
G1$\rightarrow$G3 & $209^{+2}_{-2}$ & $57^{+5}_{-8}$   & 1.16643 & $134^{+7}_{-5}$   & 0.734\\
G1$\rightarrow$G5 & $213^{+2}_{-2}$ & $52^{+7}_{-18}$  & 1.16648 & $131^{+29}_{-6}$  & 0.729\\
\hline
\end{tabular}
\tablefoot{
\tablefoottext{a}{Delensed absorber plane;}
\tablefoottext{b}{Position angle of the approaching semi-major axis (east of north);}
\tablefoottext{c}{Mass-weighted mean redshift of the considered galaxies;}
\tablefoottext{d}{Assuming a rotation curve $v(r)=v_\mathrm{rot,gas}\frac{2}{\pi}\arctan(r/r_\mathrm{t})$;}
\tablefoottext{e}{Reduced Chi-squared.}}
\end{table}

Motivated by these results, we examined a model in which the gas resides in a thick,
rotating layer or flattened spheroidal halo.
Using the \mgii\ velocities observed towards the Arc, Ci1, Ci2, and the quasar,
we applied Markov Chain Monte Carlo (MCMC) modelling with Bayesian inference to sample
the posterior distributions of position angle, inclination, and rotational velocity
within the following ranges: PA$_\mathrm{gas}=[0;360]^{\circ}$,
$i_\mathrm{gas}=[0;90]^{\circ}$, and $v_\mathrm{rot,gas}=[0;300]$~\kms. The rotation
centre and systemic redshift were allowed to differ from those of G1, considering
instead the barycentres of the two, three, or five most massive group members.
In each case, the systemic redshift was fixed to the mass-weighted mean of the
selected galaxies.
A range of vertical scale heights was also explored but found to have negligible
impact on the fits and to remain unconstrained. The best-fitting parameters are listed
in Table~\ref{tab:models_Mg}, and the model based on the barycentre and combined
redshift of G1, G2, and G3 is shown in Fig.~\ref{fig:m_model} (lower panels).
All tested rotation centres yield consistent results, indicating that the
derived parameters are robust to this choice.
This supports the presence, in addition to individual galaxy discs, of an IGrM phase
with a projected rotational velocity of
$v_\mathrm{rot,gas}\sin{i}\simeq 103$~\kms. Interestingly, the inferred rotation direction
of the halo aligns with the orbital motions of all six group galaxies (see
Fig.~\ref{fig:schematic}), suggesting that angular momentum is acquired through infall
along nearby large-scale structure filaments and subsequent mixing with
metal-enriched gas. The model also reproduces the relative velocity between G1 and G2
($\approx -94$~\kms), implying that G2 is receding from G1. If G1 and G2 have
previously interacted, G2 is now located behind G1. The spin of G1 is oriented towards
G5 and G6, consistent with filamentary accretion scenarios \citep[e.g.,][]{Welker18},
whereas the misalignment of the spin axes of G1, G2, and G3 indicates previous interactions.

Intervening \mgii\ absorption probes the entire halo, whereas gas detected in front of
individual galaxies carries the signature of their own ISM and CGM. This is particularly
evident for G3, whose DtB absorption exhibits substantial velocity dispersion (see
Table~\ref{tab:abs} and Fig.~\ref{fig:spec2}). Towards both G1 and G2, the gas is blueshifted
by $\approx -40$~\kms\ relative to the halo model, reflecting the influence of stellar winds
and/or AGN-driven outflows not included in the model. Nevertheless, the relative velocity
between these two components ($\simeq +40$~\kms) is well reproduced in both magnitude
and rotation direction, providing further support for the rotating IGrM scenario, as illustrated
in Fig.~\ref{fig:schematic}. Each spatially resolved background source (the Arc, Ci1, and Ci2; see
lower right panel of Fig.~\ref{fig:m_model}) also exhibits internal velocity scatter exceeding
that expected from pure rotation, indicating the presence of turbulence and localised gas motions
within the IGrM.

\subsection{Total gas mass}

From the observed \mgii\ gas kinematics, we can estimate the dynamical mass of the system
and constrain its total (cool$+$warm$+$hot) baryonic gas mass. To do this, we applied the
virial theorem, which describes the balance between gravitational, centrifugal, and
pressure forces for a system at equilibrium. Under this assumption, the dynamical
mass enclosed within a rotating mass distribution of radius $R$ can be written as
\citep[e.g.,][]{Epinat09,Wolf10}:
\begin{equation}\label{eq:mdyn}
M_\mathrm{dyn}(r<R)~[M_\odot] = 2.33\times 10^5\,(v_\mathrm{rot}^2 + 3\,\sigma_v^2)\,R,
\end{equation}
where $v_\mathrm{rot}$ is the rotational velocity (in \kms), $\sigma_v$ is the line-of-sight
velocity dispersion (in \kms), and $R$ is the radius of the mass distribution perpendicular
to the rotation axis (in kpc).

Strong \mgii\ absorption in the field of \sgas\ is observed out to the projected distance of
Ci2. For this reason, we considered the mass enclosed within a radius of 85~kpc, which is
the deprojected distance between Ci2 and the group's barycentre.
Using the observed mean velocity dispersion of 122~\kms\ and a rotational velocity
of 131~\kms\ (see Sects.~\ref{sec:intervening} and \ref{sec:kinematics}), we infer a total
dynamical mass of $1.22\times 10^{12}$~$M_\odot$. For comparison, the combined stellar and ISM
masses of G1, G2, and G3, enclosed within this radius, amount
to $1.52\times 10^{11}$~$M_\odot$. From dark matter theory \citep[e.g.,][]{Navarro97}, the
dark matter-to-total mass fraction at a radius of 85~kpc for a halo mass
of $1.32\times 10^{13}$~$M_\odot$ lies in the range $0.67-0.77$, depending on the
assumed parameters. This implies a total baryonic (cool$+$warm$+$hot) gas mass
of $1.3-2.5\times 10^{11}$~$M_\odot$, corresponding to approximately 55\% of all baryons
within one-quarter of the group's virial radius. When rescaled to the number of galaxies, this
corresponds to a total gas mass per galaxy of $6.3\times 10^{10}$~$M_\odot$ for a stellar
mass of $5\times 10^{10}$~$M_\odot$.

The calculation assumes the IGrM gas is in dynamical equilibrium, which may not
be strictly valid in this actively star-forming and interacting galaxy group. However, several
lines of evidence support this approximation: (1) the coherent velocity field observed
across $\ga 60$~kpc scales suggests organized rather than chaotic motions; (2) all measured gas
velocities lie well below the group escape velocity (see Fig.~\ref{fig:plot_boRvir}, right
panel), indicating gravitationally bound gas; and (3) the group's dynamical time (crossing
time) at $R\sim 85$~kpc is $\tau_\mathrm{dyn}\sim R/v_\mathrm{rot}\sim 0.65$~Gyr, sufficient
for at least partial relaxation. The virial theorem has been widely applied to estimate
masses of galaxy groups and clusters from kinematic
data \citep[e.g.,][]{Rines06,Epinat09}. Nevertheless, ongoing interactions,
turbulence, and/or magnetic pressure support likely introduce systematic uncertainties of order
50\% in our mass estimates.

From analytical modelling, \citet{Berg25} estimated the cool gas mass towards the
extended background sources of \sgas\ and derived a value of $1.2\times 10^{9}$~$M_\odot$ over
an area of 350~kpc$^2$. Within the framework of a rotating IGrM, extrapolating this mass to
a radius of 85~kpc yields a cool gas mass of $7.8\times 10^{10}\times C_f$~$M_\odot$,
where $C_f$ is the gas covering fraction. Compared to our constrained total gas mass, this
implies, for cool-gas covering fractions $C_f\gtrsim 0.5$, a cool phase accounting for
at least 20\% of the total gas mass within one-quarter of the group's virial radius, the
rest corresponding to warm$+$hot gas.

These results are consistent with previous estimates of the CGM gas mass around sub-$L^\star$
to $L^\star$ galaxies, as inferred from both observations and cosmological simulations
\citep[e.g.,][]{Werk14,Stern16,Prochaska17,Nelson20,Faerman23,Wright24}. However, unlike
isolated galaxies, the CGM of the individual galaxies in the compact group in the field of
\sgas\ appears to overlap, leading to the existence of a pervasive IGrM with its own kinematic
behaviour (see Sect.~\ref{sec:kinematics}). Based on the coherence of the observed velocity
field and the angular momentum it carries, we suggest that the cool gas in the IGrM is
kinematically locked within a warm-hot halo phase rather than reaccreting onto the parent
galaxies. The lack of reaccretion of the gas
may lead to the rapid quenching of star formation, over less than 500~Myr
\citep[e.g.,][]{Bitsakis16,Skarbinski25}, particularly in lower mass galaxies like G2.

\section{Conclusions}\label{sec:conclusions}

Gravitational-arc tomography provides a complementary approach to quasar absorption-line studies,
which probe multiphase gas in the CGM \citep{Tumlinson11,Werk14}. \citet{Chen10} found
that \mgii\ absorbers with $W^{2796}_\mathrm{r}>0.1$~\AA\ exhibit covering fractions of
approximately 80\% within 100~kpc of galaxies with luminosities $L\la L^\star$. Similarly,
\citet{Neeleman16} showed that a substantial fraction of molecular gas detected in
absorption originates from the CGM rather than the disc of an isolated galaxy at $z\simeq 0.1$.
Studies of galaxy groups indicate that absorption often arises from multiple galaxies
\citep{Bielby17,Fossati19,Dutta20,Dutta21}. While superposition models
\citep{Nielsen18,Bordoloi11,Fossati19} typically reproduce the equivalent widths, we have
shown that they do not capture the kinematics of absorption in a compact group, highlighting their
limitations in describing the IGrM. Unlike previous \arctomo\ studies, which primarily
targeted isolated galaxies \citep{Lopez18,Lopez20,Tejos21}, the present work reveals the
complexity of gas dynamics in group environments
\citep[see also][]{Cherrey24,Fernandez-Figueroa24}.

Firstly, our results demonstrate that the IGrM around star-forming galaxies in groups is abundant
in cool gas and metals.
The ubiquity and spatial coherence of the IGrM in the compact group in the field of \sgas\
suggest efficient gas processing and mixing. Although G1 was initially selected based on
prior \mgii\ absorption against the Arc, integral-field spectroscopy of all sources across
the field revealed a far greater extent and coverage of cool gas than anticipated. While cool
gas is typically expected near star-forming regions and galactic discs, our observations reveal
gas escaping in front of G1 and G2, implying metal enrichment of the halo. Moreover, the
presence of cool gas at large distances, distributed across the field, indicates that the
stripping of dwarf galaxies, in addition to star-formation-driven winds, plays a significant
role in redistributing metals. This aligns with \citet{Nielsen22}, who reported a compact galaxy
group giving rise to a DLA with ultra-strong \mgii\ absorption at $z\simeq 2.4$, and
with \citet{Puglisi21}, who observed ISM material stripped by tidal forces in a galaxy merger at
$z\simeq 1.4$.

We find evidence for a multiphase IGrM that contributes substantially to the \mgii\ absorption
signal as well as exhibits coherent large-scale rotational motions. While previous
observations have hinted at coherent motions in the IGrM \citep[e.g.,][]{Leclercq22}, we observe,
for the first time, coherent 2D kinematics of intervening absorption up to a deprojected distance
of $\sim 85$~kpc, encompassing the primary group galaxies and, most importantly, rotating
with them. Applying the virial theorem, we constrain the total baryonic (cool$+$warm$+$hot)
gas mass to $\sim 2\times 10^{11}$~$M_\odot$ within one-quarter of the group's virial
radius, with the cool phase accounting for at least 20\% of this total.
Substantial amounts of cool gas suggest ongoing growth of cool clouds within a hot halo. This
process involves interaction of cool ($\sim 10^4$~K) clouds with warm and hot gas along their
trajectories and adiabatic compression, with long dissipation timescales
\citep{Kanjilal21,Gronke18,Gronke22}.

Simulations predict that extended multiphase gas with sub-escape velocities condenses near
the wakes of satellite galaxies and intergalactic filaments on scales approaching the halo virial
radius \citep{Nelson20}.
The large-scale coherence of gas kinematics observed here supports a picture in
which metal-enriched gas is locked in a rotating halo phase, acquiring angular momentum and mixing
with gas at the intersection of infalling filaments, as predicted by cosmological
simulations \citep[e.g.,][]{Stern24}.
Our results suggest angular momentum organization across cosmic scales, from isolated
galaxy spins aligned with individual filaments \citep[e.g.,][]{Welker18,Kraljic21,Tudorache25}
to coherent group-scale rotation at their intersection.
Consequently, intervening absorbers may not always be associated with individual
galaxies, particularly in massive halos \citep[e.g.,][]{Weng23}.
Instead, high-redshift strong \mgii\ absorbers may trace shared, metal-enriched halos shaped by
galaxy interactions and feedback, with stripped and outflowing gas accumulating in the IGrM
over time.

\begin{acknowledgements}

We are grateful to Andrea Afruni and Rongmon Bordoloi for insightful discussions. CL
acknowledges
support from ESO during his research period. FM and FB acknowledge funding from ANID through
Becas/Mag\'\i ster Nacional 22240417, Basal project FB210003, FONDECYT grant 1230231,
and support from CATA and ESO.
NT acknowledges support from FONDECYT grant 1231187.
This work is based on observational data collected by the \arctomo\ collaboration
(https://sites.google.com/view/arctomo/home).

\end{acknowledgements}

\bibliographystyle{aa}
\bibliography{aa58070-25}

\begin{appendix}

\section{Galactic foreground contamination}

\begin{figure}
\begin{minipage}{1.\columnwidth}
\includegraphics[width=\columnwidth]{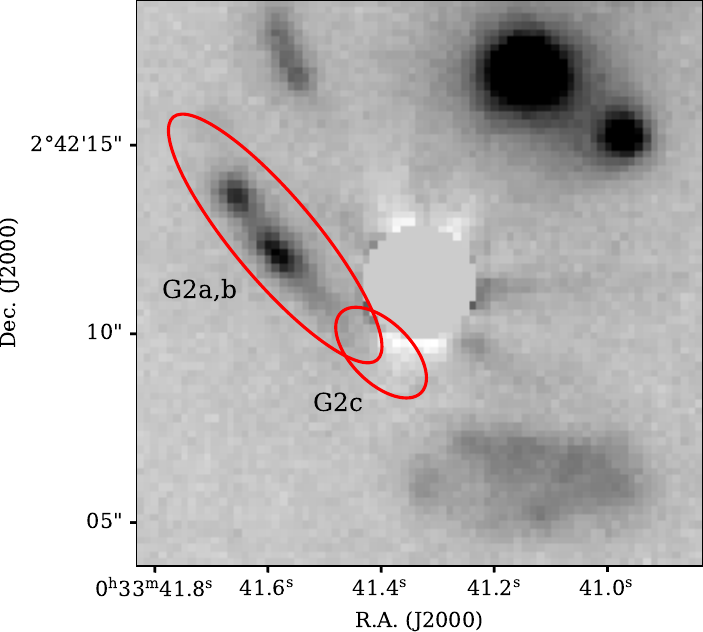}
\end{minipage}
\caption{PSF subtraction and aperture masking of the Galactic star located between G1 (southwest) and G2 (east) in a
MUSE narrow-band
image centred on \oii\ emission at $z=1.17$.
Light in this image may also arise from continuum emission. G2a and G2b refer to the brightest knots
within G2. G2c denotes the region of G2 nearest to the stellar core (masked out) where \oii\ emission can be detected.}
\label{fig:psfsub}
\end{figure}

A bright Galactic star located near the centre of the field (see Fig.~\ref{fig:field}) is 
posing a challenge for the study of the region between G1 and G2 due to light contamination. This
star, catalogued as GAIA DR3 2553567450013494784, is a K2V-type star with a mean G-band
magnitude of 14.8 \citep{Gaia23}. To mitigate its impact on our targets, we adopted the following
procedure. First, we modelled the point spread function (PSF) using three faint stars located
in uncrowded regions of the MUSE white-light image, fitting 2D Moffat profiles. The resulting
fits were consistent with one another, yielding best-fit parameters
of FWHM$=0\farcs 86$ at $8000$~\AA\ and $\beta=2.14$, in agreement with the results of
\citet{Bacon23}.

Given its brightness, the central star exhibits artefacts caused by internal reflections
and flat-fielding errors, which are apparent in the MUSE white-light image. For this reason,
we focused on a narrow-band image centred on the \oii\ emission lines of interest
(see Sect.~\ref{sec:census}). We masked the stellar core using a circular aperture with a
radius of eight spaxels, then fitted a 2D Moffat profile to the outer regions. This model was
subtracted from the original image to produce a corrected version. While this method improved
the representation of the annular region beyond an angular separation of $1\farcs 6$ from the
stellar core, residuals persisted in the innermost region. Consequently, this central region
was masked and excluded from further analysis.

Fig.~\ref{fig:psfsub} shows the residuals around the star after PSF subtraction and aperture
masking. This cleaned image was used to measure star-formation rates based on \oii\ emission
(see Sect.~\ref{sec:sfr}).

\section{Magnification map}

\begin{figure}
\begin{minipage}{1.\columnwidth}
\includegraphics[width=\columnwidth]{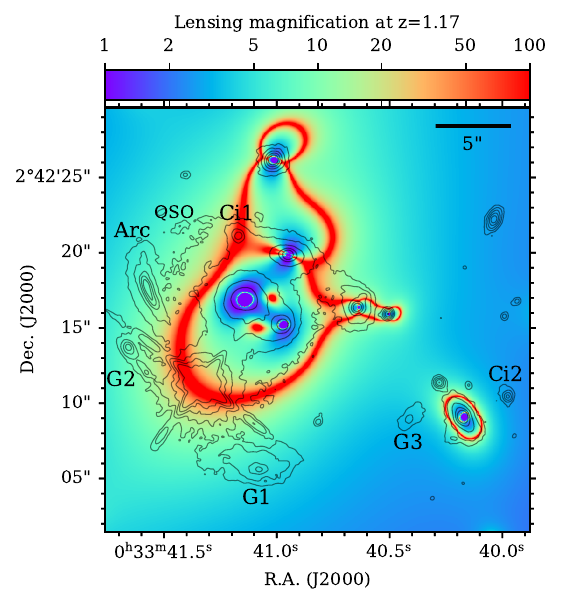}
\end{minipage}
\caption{Magnification map of sources at $z=1.17$, with critical lines highlighted in red. Black contours represent
the \hst/F140W image, overlaid to show the magnification of G1, G2, and G3, which are located at the same redshift
as the detected \mgii\ absorption.}
\label{fig:magni}
\end{figure}

In this appendix, we present the magnification map of the lensing model
(see Fig.~\ref{fig:magni}).

\section{Emission and absorption lines}\label{sec:intspec}

\begin{figure*}
\includegraphics[width=0.635\columnwidth]{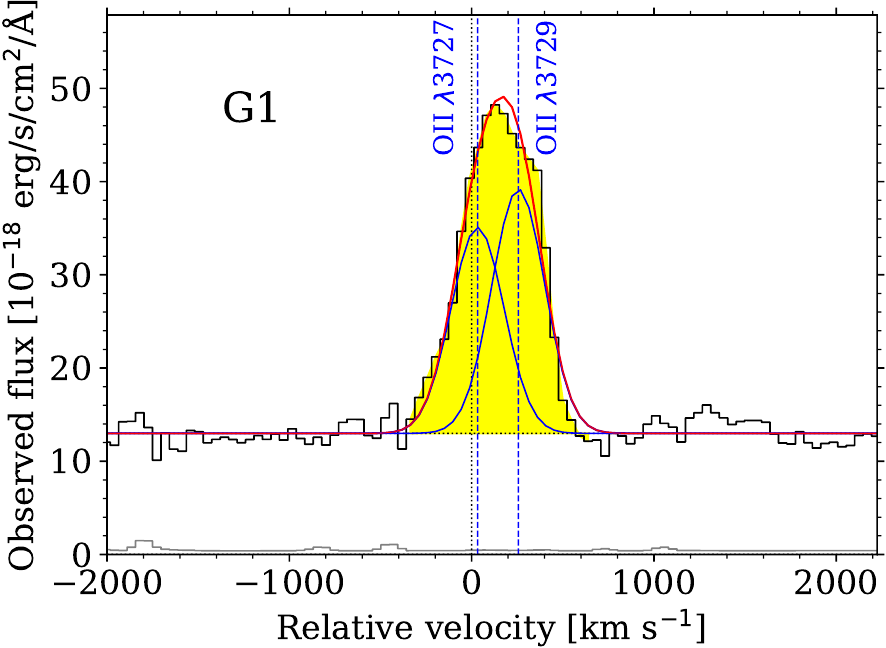}
\includegraphics[width=0.635\columnwidth]{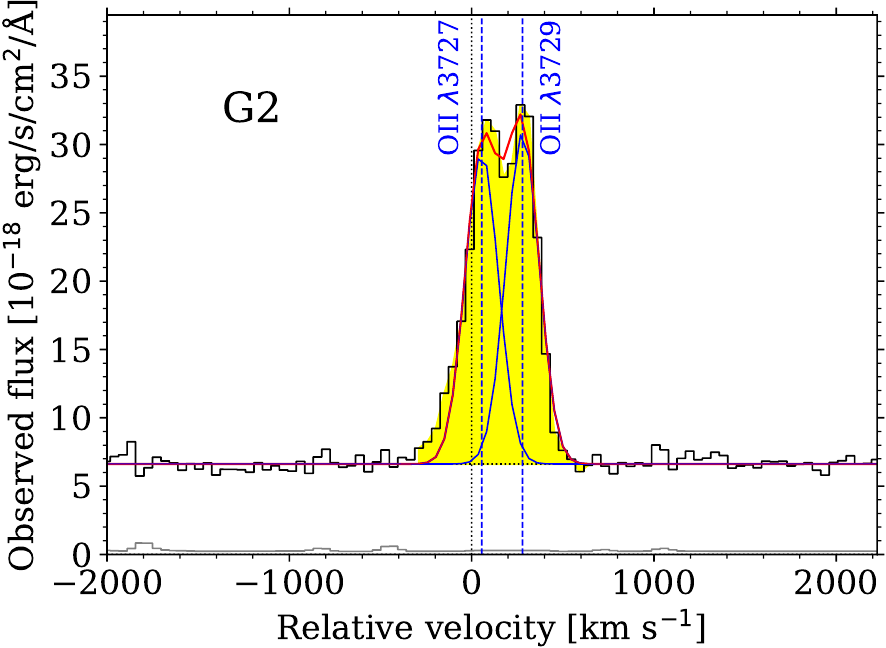}
\includegraphics[width=0.635\columnwidth]{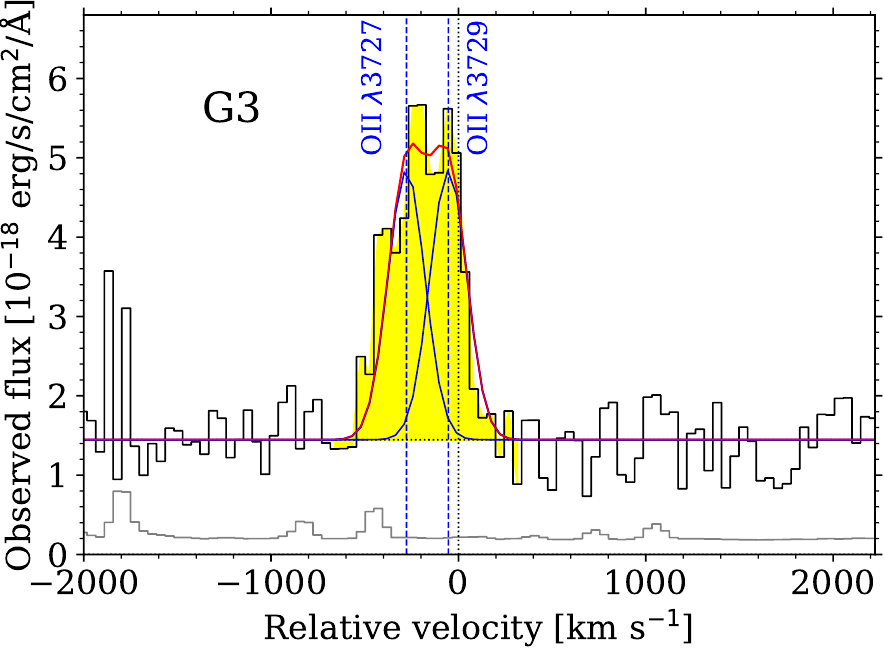}\\
\includegraphics[width=0.635\columnwidth]{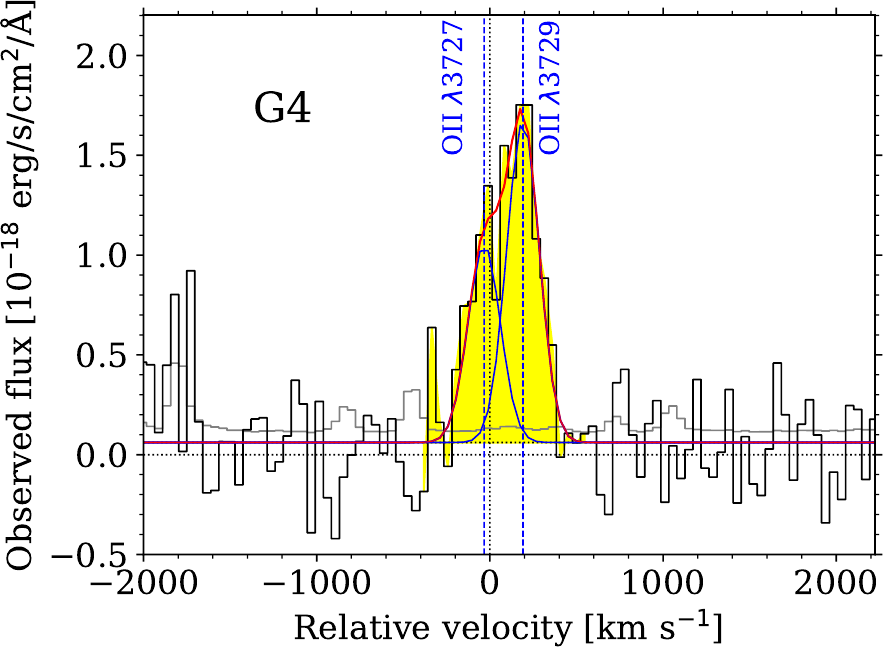}
\includegraphics[width=0.635\columnwidth]{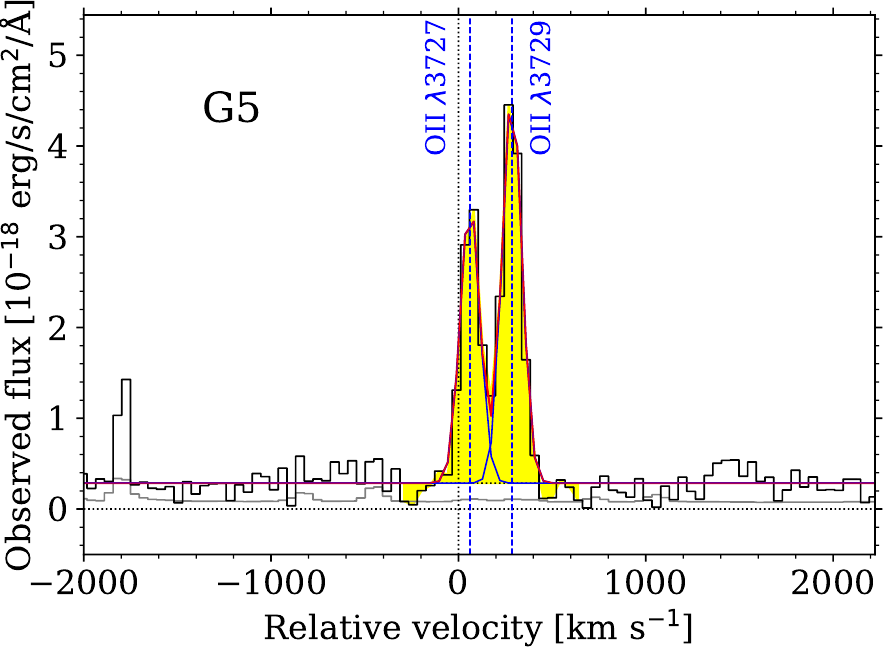}
\includegraphics[width=0.635\columnwidth]{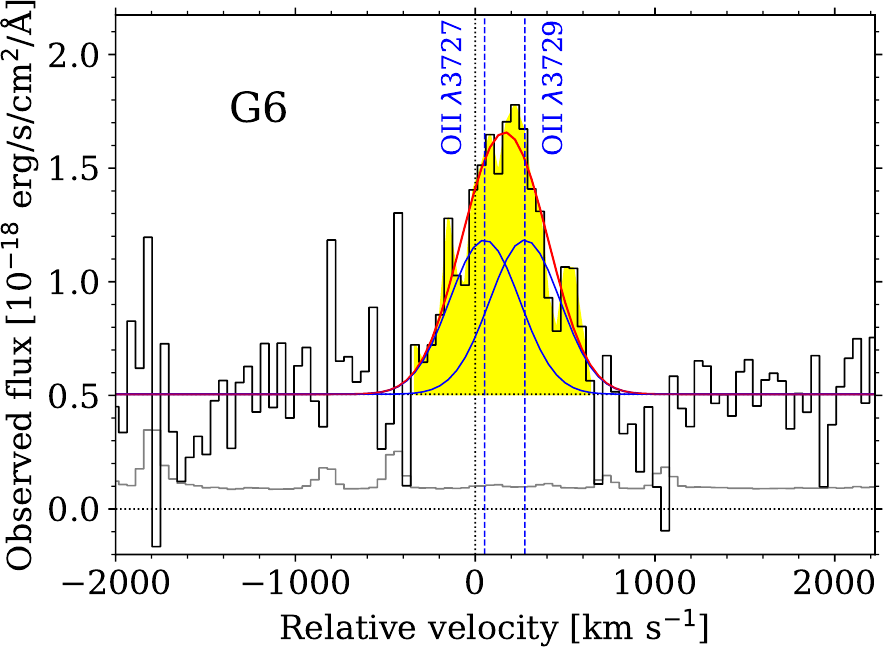}
\caption{Nebular \oii\ emission in individual galaxies, with binned fluxes and $1\sigma$
uncertainties shown in black and grey, emission lines in yellow, and double-Gaussian fits in red
and blue. The zero point of the velocity scale corresponds to $z=1.167$.}
\label{fig:spec1}
\end{figure*}

\begin{figure*}
\includegraphics[width=0.635\columnwidth]{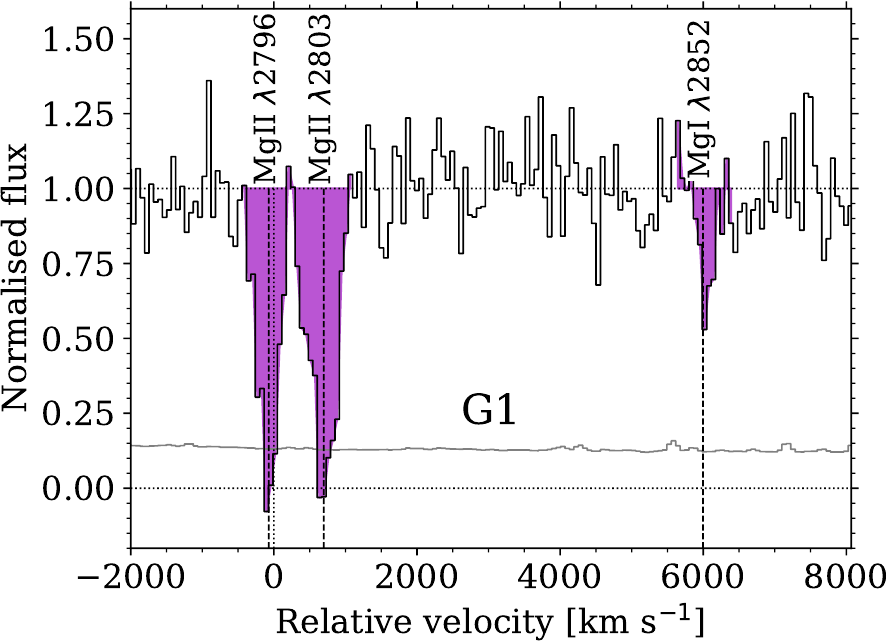}
\includegraphics[width=0.635\columnwidth]{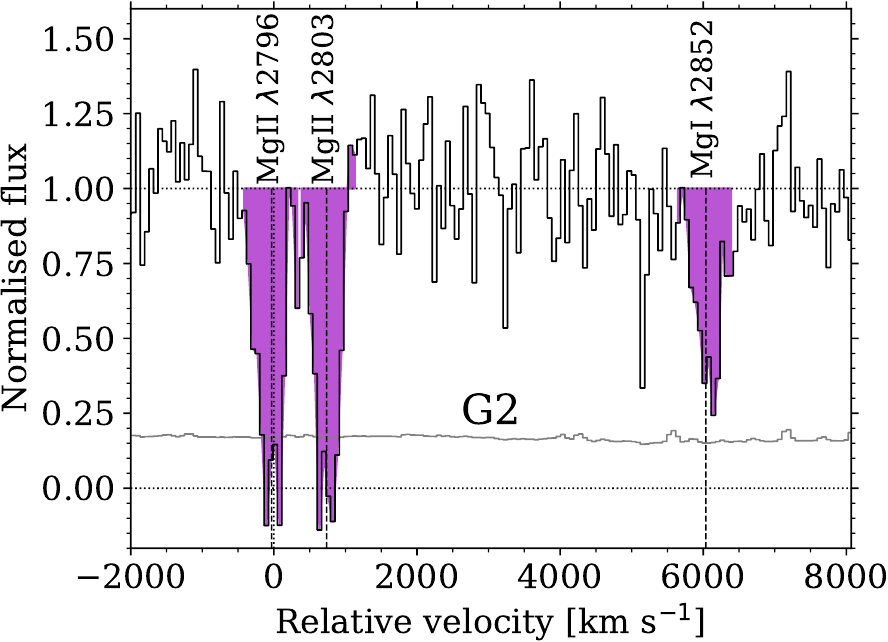}
\includegraphics[width=0.635\columnwidth]{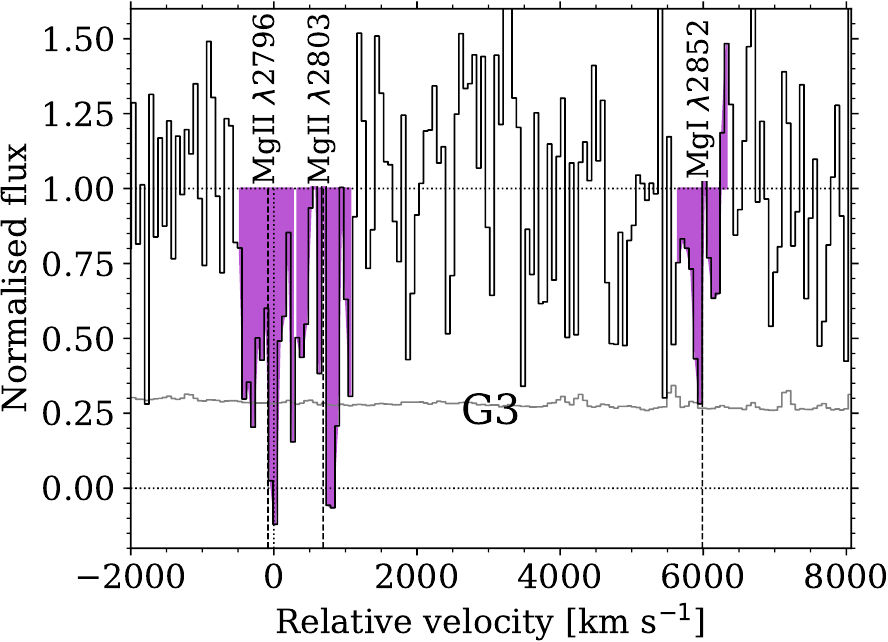}\\
\includegraphics[width=0.635\columnwidth]{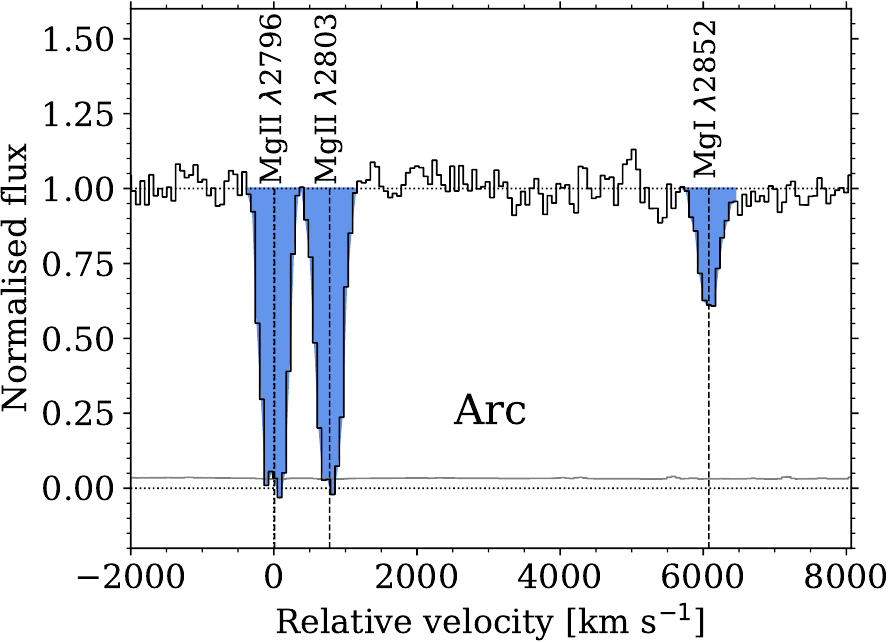}
\includegraphics[width=0.635\columnwidth]{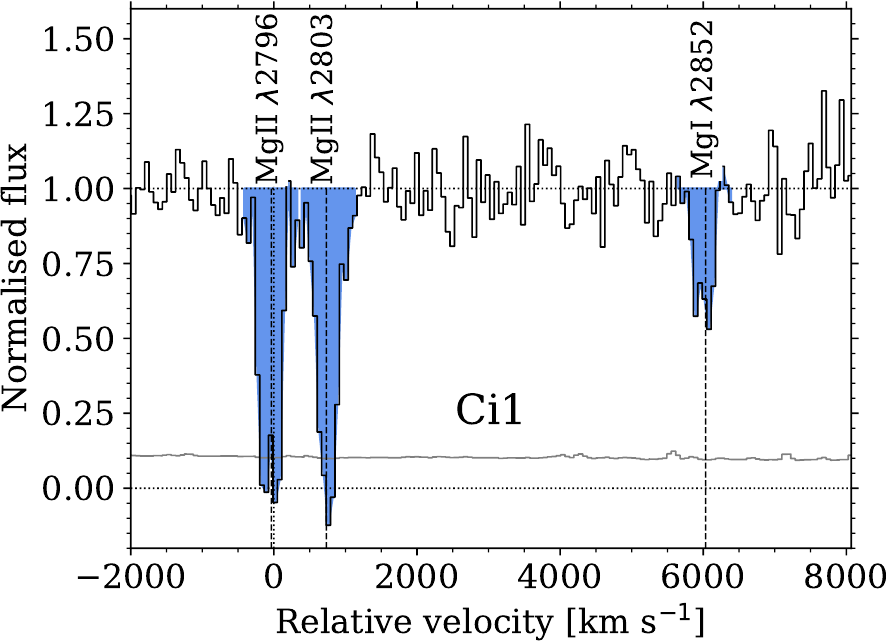}
\includegraphics[width=0.635\columnwidth]{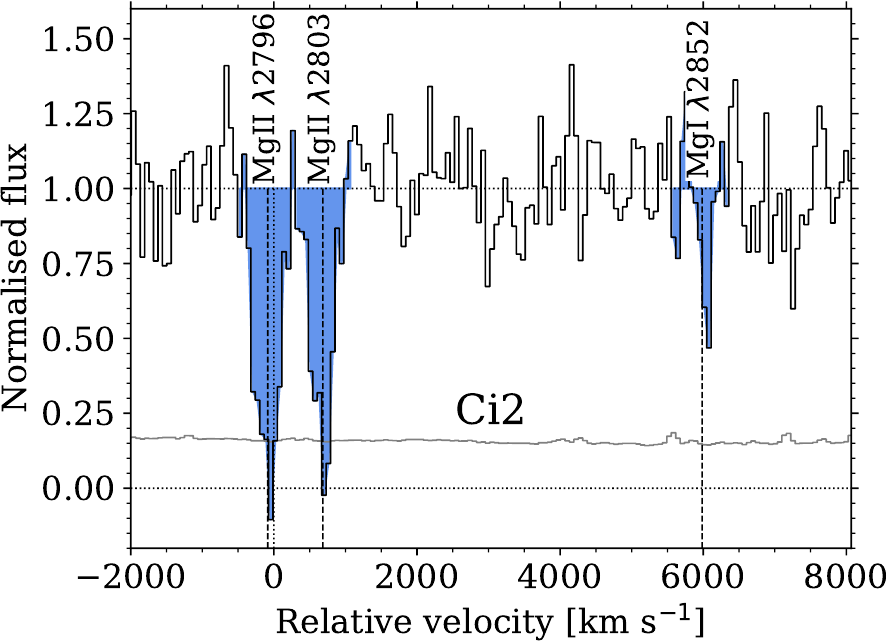}\\
\includegraphics[width=0.635\columnwidth]{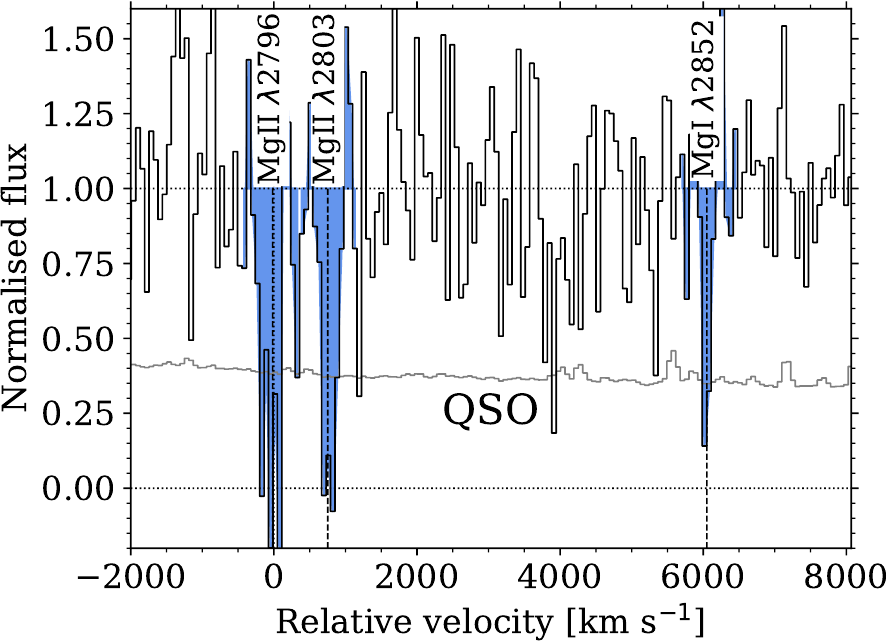}
\caption{Down-the-barrel \mgi\ and \mgii\ absorption towards G1, G2, and G3 (top panels), and
intervening absorption towards the Arc, Ci1, Ci2, and the quasar (middle and bottom panels).
Normalised binned fluxes and $1\sigma$ uncertainties are shown in black and grey, with
absorption lines filled in purple (top) or blue (middle and bottom). The zero point of the velocity
scale corresponds to $z=1.167$.}
\label{fig:spec2}
\end{figure*}

Nebular \oii\ emission and \mgii\ and \mgi\ absorption lines detected at $z\simeq 1.17$
in the integrated MUSE spectra are shown in Figs.~\ref{fig:spec1} and \ref{fig:spec2}
respectively.

\section{Intensity and velocity maps}\label{sec:maps}

\begin{figure*}
\includegraphics[width=\hsize]{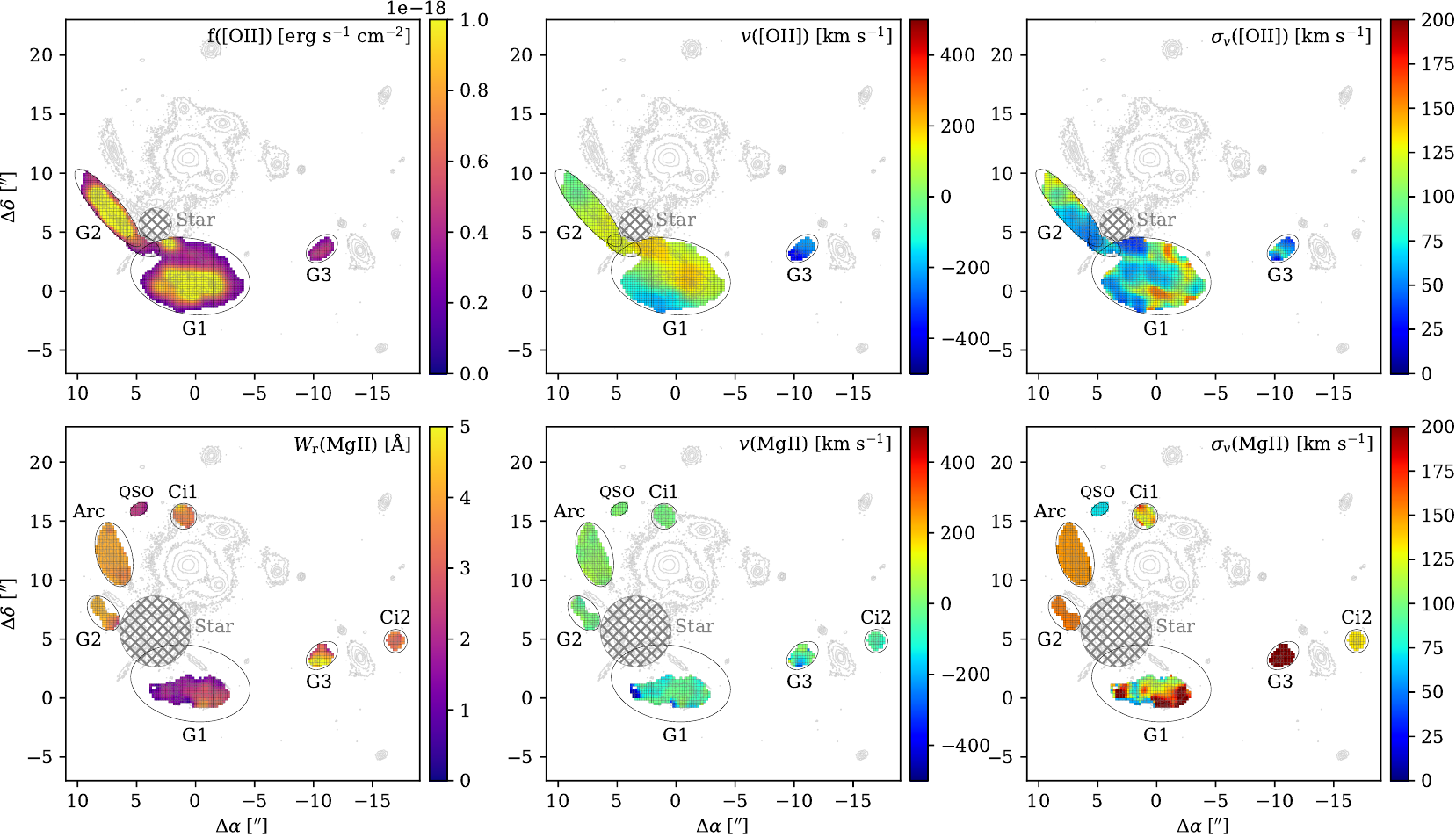}
\caption{Maps of line fluxes (or equivalent widths), line-of-sight velocities, and velocity dispersion
of the nebular
\oii\ emission (top panels) and \mgii\ absorption (bottom panels) near the group centre, in the
image plane. The orientation is north up and east to the left. The colour bar on the right-hand side
of the plots indicates the measured quantity in the units specified in the inset. The zero point of
the velocity scale corresponds to $z=1.167$.}
\label{fig:velmaps_imgplane}
\end{figure*}

\begin{figure*}
\includegraphics[width=\hsize]{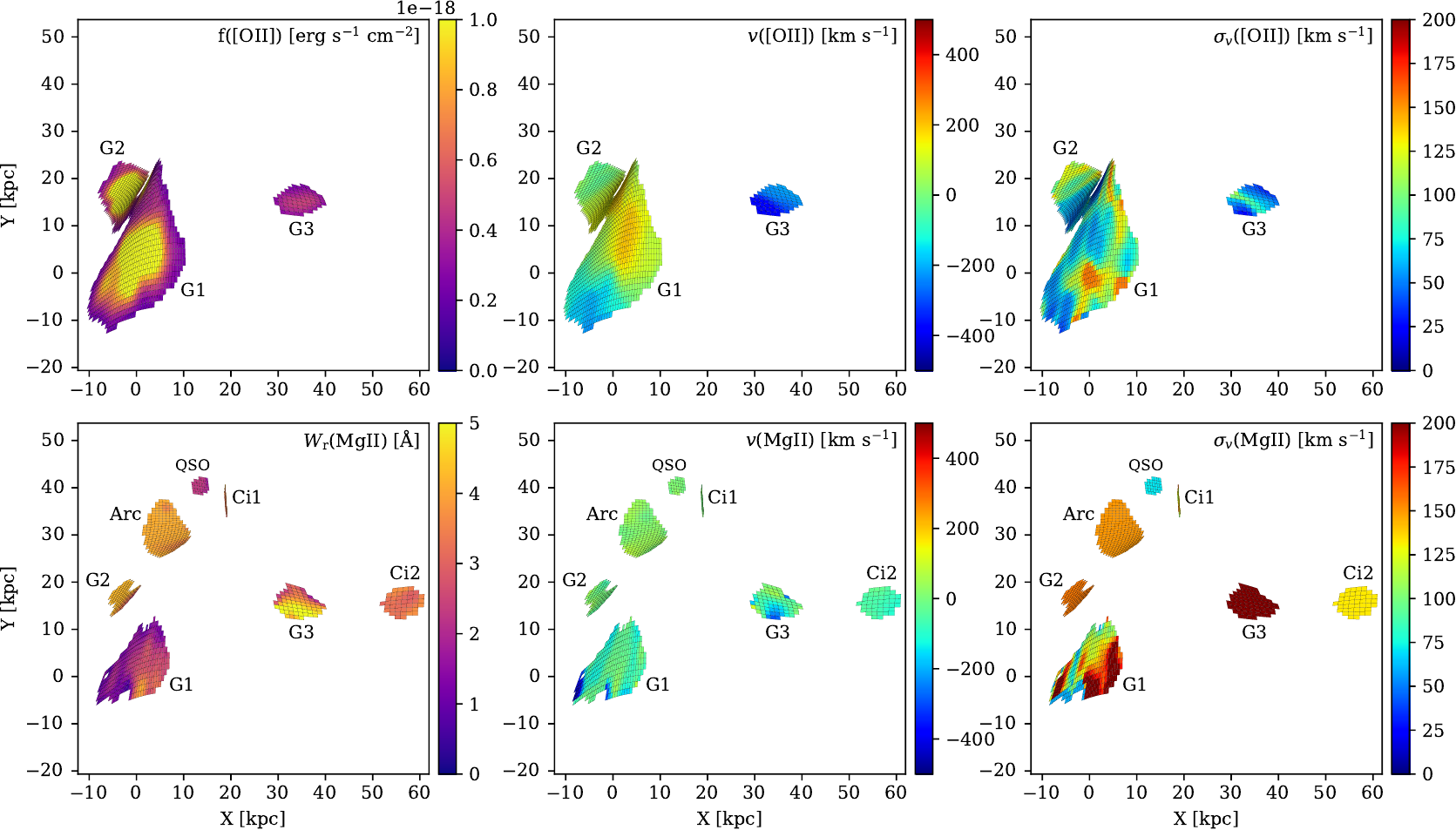}
\caption{Same as Fig.~\ref{fig:velmaps_imgplane}, but in the delensed absorber plane.}
\label{fig:velmaps_absplane}
\end{figure*}

To derive an accurate representation of the data,
we used the MUSE datacube to extract spatially-resolved information across the field at the redshift of the
absorbing galaxy group.
We employed a method similar to the one described in \citet{Berg25} but with notable exceptions.
We utilised
the MUSE Python Data Analysis Framework, \mpdaf\ \citep{Piqueras19}, and first applied Gaussian resampling
to the datacube
in the spatial direction with $\sigma=1.7$ spaxel, matching the seeing conditions during
the observations ($\mathrm{FWHM}_\mathrm{IQ}\simeq 0\farcs 8$). This way, the information contained in each spaxel and the S/N
ratio is optimised while preserving the native spatial resolution of the dataset.
Each object of interest -- G1, G2, G3, the Arc, Ci1, Ci2, and the quasar -- were singled out in the image plane
using elliptical apertures tailored to this resampled datacube, analogue to the method
we employed to analyse the \hst\ images (see Fig.~\ref{fig:delens}).
In addition, the central glare of the Galactic star, extending to a minimum diameter of $1\farcs 4$, prevents any useful
information about \oii\ and \mgii\ to be extracted, even after subtraction of the stellar PSF,
and was masked out.

Spectral continuum fitting was performed using \mpdaf\ on each spaxel over a velocity range
of 1000~\kms\ around the emission and/or absorption lines of interest using zero-order polynomial functions. We
found that this setup provided an optimal balance between accuracy and robustness of the results. The contribution of
the sky background to the signal was estimated within concentric elliptical apertures around each object and
subtracted from the signal measured in each spaxel.

The significance of the object's signal in each spaxel was assessed based on either the S/N ratio of the spectrum in
the continuum adjacent to the line or the total S/N ratio of the spectrum in the searched window after
continuum subtraction, depending on whether the feature to be mapped was an absorption or an emission line, respectively.
For absorption lines, we focussed on the \mgii/\mgi\ lines near 2800~\AA\ (rest frame) and set a
minimum S/N ratio of 2.0 per pixel in the dispersion direction to ensure that the object's continuum is bright
enough to constrain the absorption lines (if any). This threshold corresponds to the detection of the three lines combined
at the $5\sigma$ confidence level. The chosen parameters also enabled us to detect the
\mgii\ signal on top of the quasar continuum, in spite of its faintness. For emission lines, we
used a minimum S/N ratio of 1.0 per pixel in the dispersion direction
around [\ion{O}{ii}] after continuum subtraction, which translates to a $4\sigma$ detection limit for a line spread over
15-20 pixels. Spaxels that did not meet these criteria were not considered any further.

We utilised \pyspeckit\ \citep{Ginsburg22} for line fitting. To ensure robustness, we employed double
Gaussian profiles with identical
redshifts and spectral line widths to model the
\oii\,$\lambda\lambda$3727,3729 emission line doublet. Likewise, we used a single Gaussian profile for each
of the three Mg lines with fixed wavelength ratios and identical line widths. In instances of non-detection,
we established upper limits on rest-frame equivalent widths based on the root mean square (RMS) of the noise in
the adjacent continuum.
The resulting 2D maps of equivalent widths, line-of-sight velocities, and velocity dispersion, are shown in the
image plane in Fig.~\ref{fig:velmaps_imgplane}.
We then applied the deflection matrices (see Sect.~\ref{sec:lens}) to the positions of the spaxel corners (i.e.,
vertices) in the image plane to recover their delensed positions in the absorber plane. This leads to the maps shown
in Fig.~\ref{fig:velmaps_absplane}.

The spatial resolution of these maps is constrained by the image quality, and this effect is not corrected before
transforming the object geometries to the absorber plane. As a result, G2 appears more elongated in the southeastern-northwestern
direction in the delensed maps than it is in reality.
Additionally, due to the geometrical effects of lensing, Ci1 samples a narrow range in R.A. within the absorber plane,
forming a slender, north-south elongated structure that is hardly visible on the absorption-line maps.
The region of maximum velocity dispersion in G1’s core can be used to derive the coordinates of G1's centre in the
absorber plane, i.e.,
R.A. (J2000) $=0^\mathrm{h}$ $33^\mathrm{m}$ $40\fs 91$, Dec. (J2000) $=+2\degr$ $42\arcmin$ $22\farcs 3$. The
distribution of velocities along G1’s major axis is found to turn around, changing from negative to positive values, at
this position for $z_\mathrm{sys}=1.16677$.

Overall, we find no indication that the velocity measurements derived from emission or absorption lines are
affected by lower S/N ratios. On the other hand, the determinations of gas velocity dispersion (i.e., line widths)
exhibit considerable scatter, and their reliability is likely diminished in spaxels with weak
absorption and/or noisy spectra.

\begin{figure*}
\includegraphics[width=\hsize]{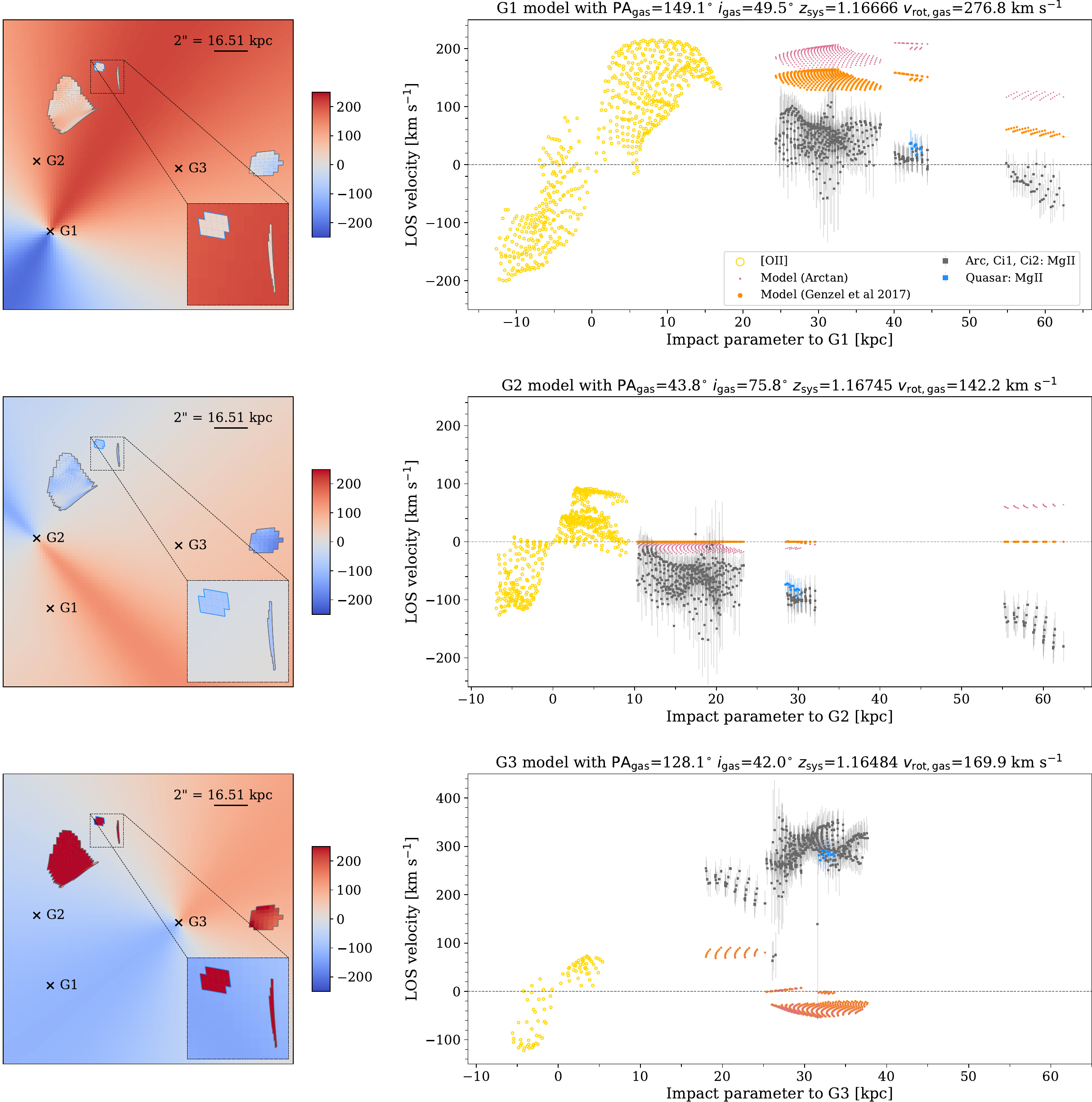}
\caption{Single extended rotating disc models for G1, G2, and G3 (top, middle, and bottom panels,
respectively) based on extrapolations of their \oii\ kinematics. {\it Left panels:} Modelled
velocity maps assuming arctangent rotation curves. The observed
intervening \mgii\ mean velocities are overlaid for each of the four delensed sources. The inset shows
a zoom-in on a region encompassing Ci1 and the quasar. {\it Right
panels:} Model-predicted velocities for both the arctangent and realistic rotation curves
(in pink and
orange, respectively) at the positions of the Arc, Ci1, Ci2, and the quasar, as a function of impact
parameter relative to the galaxy. The observed \mgii\ velocities are displayed in black for the
Arc, Ci1, and Ci2, and in blue for the quasar.}
\label{fig:g_model}
\end{figure*}

\end{appendix}

\end{document}